\documentclass[reprint,NumberedRefs]{JASA}

\usepackage{algpseudocode}
\usepackage{mathrsfs}
\usepackage{mathtools}

\begin{document}

\title[Iijima \textit{et al.}]{Binaural rendering from microphone array signals of arbitrary geometry}
\author{Naoto Iijima}
\author{Shoichi Koyama}
\author{Hiroshi Saruwatari}
\affiliation{The University of Tokyo, Graduate School of Information Science and Technology, 7-3-1 Hongo, Bunkyo-ku, Tokyo 113-8656, Japan}



\begin{abstract}
A method of binaural rendering from microphone array signals of arbitrary geometry is proposed.
To reproduce binaural signals from microphone array recordings at a remote location, a spherical microphone array is generally used for capturing a soundfield. However, owing to the lack of flexibility in the microphone arrangement, the single spherical array is sometimes impractical for estimating a large region of a soundfield. We propose a method based on harmonic analysis of infinite order, which allows the use of arbitrarily placed microphones. In the synthesis of the estimated soundfield, a spherical-wave-decomposition-based binaural rendering is also formulated to take into consideration the distance in measuring head-related transfer functions. We develop and evaluate a composite microphone array consisting of multiple small arrays. Experimental results including those of listening tests indicate that our proposed method is robust against change in listening position in the recording area.
\end{abstract}


\newcommand{\sumsum}{\sum_{n=0}^{\infty}\sum_{m=-n}^n}
\newcommand{\argmax}{\mathop{\rm arg~max}\limits}
\setlength\floatsep{5pt}
\setlength\textfloatsep{5pt} 
\setlength\intextsep{5pt}
\setlength\abovecaptionskip{5pt}

\maketitle

\section{\label{sec:1} Introduction}

Spatial sound reproduction is essential to achieve virtual or augmented reality (VR/AR) audio systems; therefore, a wide range of spatial audio techniques have been developed. Soundfield reproduction/control using multiple loudspeakers makes it possible to reproduce a spatial sound inside a relatively large region~\cite{Berkhout:JASA_J_1993,Daniel:AES114conv,Poletti:J_AES_2005,Spors:AES124conv,Koyama:IEEE_J_ASLP2013}. On the other hand, binaural rendering techniques reproduce sounds received in both the listener's left and right ears generally through headphones. Binaural signals are typically synthesized using transfer functions from sources to the eardrums in a free field, i.e., the so-called \textit{head-related transfer functions (HRTFs)}, measured in advance. An HRTF includes the effects of diffraction and reflection caused by the listener's head, torso, and external pinna shape, which contains important features for sound localization in humans~\cite{HRTF}. 

When virtual source positions are given, it is possible to synthesize the binaural signals in a free field by the direct convolution of transfer functions and source signals; however, it is not easy to synthesize binaural signals from signals captured by microphones in a reverberant environment, where the synthesized signals should correspond to the signals received in both ears when the listener is present in the recording area. It is also preferable that the binaural signals vary in response to the listener's head movement from the perspective of the sound localization property in humans~\cite{headmove}. One of the solutions will be the use of a dummy head that moves synchronously with the listener's movement~\cite{telehead}. However, it is difficult to simultaneously bring the binaural signals to multiple listeners. One of the promising approaches is binaural reproduction from microphone array recordings~\cite{senzi,vah,vah2,beam,beam2,beam3,conv2}. Owing to the recent developments of spatial soundfield recording techniques, it is becoming possible to estimate a soundfield inside a region using multiple microphones~\cite{Park:JASA_J_2005,Poletti:J_AES_2005,micarray,samarashinge_2014_ASLP,Ueno:IEEE_SPL2018,Nakanishi:WASPAA2019}. The reproduced binaural signals can be highly accurate in a physical sense at low frequencies owing to the use of the spatial soundfield information estimated by these techniques. 

A typical microphone array used in the soundfield recording is a spherical microphone array mounted on an acoustically rigid object~\cite{conv2} since the spherical harmonic analysis of a soundfield is effective for binaural rendering~\cite{conv,plane_trans,schultz2013data-based,hrtf_pw_modal,hrtf_sph_prepro2}. The number of microphones and array size limit the reproducible frequency range and size of the reproducing region. In particular, for the binaural reproduction, the region of high estimation accuracy should be larger than the head size. 
Therefore, the use of a single spherical microphone array is sometimes impractical owing to the lack of flexibility in the microphone arrangement. 

We propose a method of binaural reproduction from distributed microphone array signals. The soundfield estimation method based on harmonic analysis of infinite order is applied to obtain harmonic coefficients of the soundfield in the recording area. This estimation method makes it possible to use arbitrarily placed microphones for capturing a sound field. For example, multiple small microphone arrays, such as low-order spherical microphones (ambisonics microphones) and acoustic vector sensors, will be more practically feasible than a single large spherical microphone array. Moreover, such a composite array will have flexibility and scalability in the array placement. In the binaural rendering from the estimated harmonic coefficients, we use the soundfield representation by spherical wave decomposition~\cite{Iijima:IEEE_MMSP2020}. In contrast to a typically used plane wave decomposition method, our rendering method takes into consideration the distance in HRTF measurement. In numerical simulations, it is shown that the binaural signals can be accurately reproduced by using our soundfield estimation method with the composite array of microphones. Moreover, the peaks and dips in the amplitude response of the binaural signals reproduced on the basis of the conventional plane wave decomposition method can be shifted from the true ones. They are accurately reproduced by our spherical-wave-decomposition-based binaural rendering method. We also develop a practical microphone array consisting of multiple small microphone arrays, and the binaural signals reproduced by using this array system are also evaluated. 

The rest of this paper is organized as follows. In Sect.~\ref{sec:preliminaries}, we introduce several notations and basic theories on sound field representation. The problem statement and prior works are introduced in Sect.~\ref{sec:bnrl_pwd}. In Sect.~\ref{sec:pro}, we describe our proposed binaural rendering method for distributed microphone array signals. Numerical simulation results for comparing binaural rendering and soundfield estimation methods are shown in Sects.~\ref{sec:pln_vs_sph} and \ref{sec:distmic}, respectively. Experimental results in a practical environment are shown in Sect.~\ref{sec:3}. Finally, our conclusion is presented in Sect.~\ref{sec:conclusion}.

\section{\label{sec:preliminaries} Preliminaries}

First, we introduce several notations and basic theories used throughout this paper. 

\subsection{\label{subsec:notations} Notations}

Italic letters denote scalars, lower case italic boldface letters denote vectors, and upper case italic boldface letters denote matrices. The sets of real and complex numbers are denoted by $\mathbb{R}$ and $\mathbb{C}$, respectively. The unit sphere in $\mathbb{R}^3$ is denoted by $\mathbb{S}_2$. Subscripts of scalars, vectors, and matrices indicate their indexes. For example, $x_{i,j}$ is the $(i,j)$th entry of matrix $\bm{X}$. 

The imaginary unit is denoted as $\mathrm{j}=\sqrt{-1}$. The complex conjugate, transpose, conjugate transpose, and inverse are denoted by superscripts $(\cdot)^{\ast}$, $(\cdot)^{\mathsf{T}}$, $(\cdot)^{\mathsf{H}}$, and $(\cdot)^{-1}$, respectively. The absolute value of a scalar $x$ and the Euclid norm of a vector $\bm{x}$ are denoted as $|x|$ and $\|\bm{x}\|$, respectively. The inner product of vectors $\bm{x}$ and $\bm{y}$ is denoted as $\langle \bm{x}, \bm{y} \rangle$. 

The functions $j_n(\cdot)$ and $h_n(\cdot)$ are the $n$th order spherical Bessel function and $n$th-order spherical Hankel function of the second kind, respectively. The spherical harmonic function of order $n$ and degree $m$ for azimuth and zenith angles ($\phi$ and $\theta$, respectively) is defined as
\begin{align}
Y_n^m(\theta,\phi) = \sqrt{\frac{2n+1}{4\pi}\frac{(n-m)!}{(n+m)!}}P_n^m(\cos\theta) e^{\mathrm{j}m\phi},
\end{align}
where $P_n^m(\cdot)$ is the associated Legendre function of order $n$ and degree $m$. 

\subsection{\label{subsec:sound} Sound field representation using spherical wavefunctions}

The stationary soundfield of angular frequency $\omega$ inside a region not including any sources $D$ satisfies the homogeneous Helmholtz equation as
\begin{align}
 (\Delta + k^2) u(\bm{r},k) = 0,
\label{eq:helmeq}
\end{align}
where $\Delta$ is the Laplacian and $u(\bm{r},k)$ is the sound pressure at the position $\bm{r} \in D$ and wavenumber $k$ defined as $k=\omega/v$ with sound velocity $v$. The solution of Eq.~\eqref{eq:helmeq} for the source-free interior region $D$ can be expanded using spherical wavefunctions at the expansion center $\bm{r}_{\mathrm{c}}$ as
\begin{align}
 u(\bm{r},k) = \sum_{n=0}^{\infty} \sum_{m=-n}^n \alpha_{n}^m(\bm{r}_{\mathrm{c}}, k) \varphi_n^m(\bm{r}-\bm{r}_{\mathrm{c}},k).
\label{eq:shd}
\end{align}
Here, $\varphi_n^m(\cdot)$ is the spherical wavefunction defined for $\bm{r}=[r, \theta, \phi]^{\mathsf{T}}$ in the spherical coordinates as
\begin{align}
 \varphi_n^m(\bm{r},k) = \sqrt{4\pi} j_n(kr) Y_n^m(\theta,\phi).
\end{align}
We also represent Eq.~$\eqref{eq:shd}$ in a matrix form as
\begin{align}
 u(\bm{r},k) = \bm{\alpha}(\bm{r}_{\mathrm{c}},k)^{\mathsf{T}} \bm{\varphi}(\bm{r}-\bm{r}_{\mathrm{c}},k),
\end{align}
where $\bm{\alpha}(\bm{r}_{\mathrm{c}},k)\in\mathbb{C}^{\infty}$ and $\bm{\varphi}(\bm{r}-\bm{r}_{\mathrm{c}},k)\in\mathbb{C}^{\infty}$ are the infinite dimensional vectors of the expansion coefficients $\alpha_{n}^m(\bm{r}_{\mathrm{c}},k)$ and spherical wavefunctions $\varphi_{n}^m(\bm{r}-\bm{r}_{\mathrm{c}},k)$, respectively. Note that $\alpha_0^0(\bm{r},k)$ corresponds to $u(\bm{r},k)$. 

The expansion coefficients at two expansion centers $\bm{r}$ and $\bm{r}^{\prime}$ ($\bm{\alpha}(\bm{r},k)$ and $\bm{\alpha}(\bm{r}^{\prime},k)$, respectively) can be related as
\begin{align}
\bm{\alpha}(\bm{r},k) = \bm{T}(\bm{r}-\bm{r}^{\prime},k)\bm{\alpha}(\bm{r}^{\prime},k),
\label{eq:trans_coef}
\end{align}
where $\bm{T}(\bm{r}-\bm{r}^{\prime},k)\in\mathbb{C}^{\infty\times\infty}$ is the translation operator defined as 
\begin{align}
\left[ \bm{T}(\bm{r})\bm{\alpha}\right]_{n'}^{m'} &= \sumsum [ 4\pi (-1)^m \mathrm{j}^{n'-n} \nonumber \\
& \hspace{-1cm}  \cdot \sum_{l=0}^{n'+n} \mathrm{j}^l j_l(kr)Y_l^{m'-m}(\theta,\phi)^* \mathcal{G}(n,m;n',-m';l) ] \alpha_n^m.
\label{eq:def_trans}
\end{align}
Here, $[\cdot]_{n^{\prime}}^{m^{\prime}}$ is the element of order $n^{\prime}$ and degree $m^{\prime}$, and $\mathcal{G}(\cdot)$ is the Gaunt coefficient~\cite{multiple}. The translation operator satisfies the following equations:
\begin{align}
\bm{T}(-\bm{r},k) &= \bm{T}^{-1}(\bm{r},k) = \bm{T}^\mathsf{H}(\bm{r},k), \\
\bm{T}(\bm{r}+\bm{r}^{\prime},k) &= \bm{T}(\bm{r},k)\bm{T}(\bm{r}^{\prime},k).
\label{eq:trans_prop}
\end{align}

\subsection{\label{subsec:hrtf} Head-related transfer function and its representation in spherical harmonic domain}

\begin{figure}[t]
\begin{center}
\includegraphics[width=5.0cm]{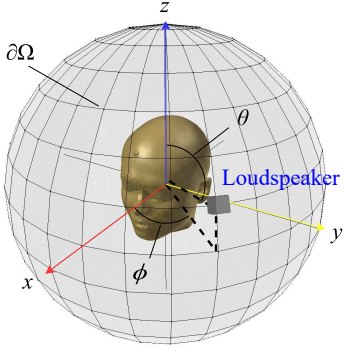}
\caption{Geometry for HRTF measurement in spherical coordinates. The transfer functions from the loudspeaker on a spherical surface $\partial\Omega$ to the listener's left and right eardrums are obtained.}
\label{fig:geometry}
\end{center}
\end{figure}

Frequency-domain transfer functions from sources to the listener's left and right eardrums are usually measured using loudspeakers on a spherical surface in an anechoic chamber~\cite{datahrtf2}. We refer to these transfer functions as HRTFs although an HRTF is generally defined as that divided by the transfer function from the loudspeaker to the center of the head without considering the listener. As shown in Fig.~\ref{fig:geometry}, we assume that HRTFs are measured from loudspeakers on a spherical surface $\partial \Omega$, whose radius is denoted as $R_{\mathrm{s}}$, in a free field. The HRTFs for the left and right eardrums at the wavenumber $k$ are denoted as $h_{\mathrm{L}}(\bm{r}_{\mathrm{s}},k)$ and $h_{\mathrm{R}}(\bm{r}_{\mathrm{s}},k)$, respectively, with the position of the loudspeaker $\bm{r}_{\mathrm{s}}\in\partial\Omega$. Hereafter, the subscripts $\mathrm{L}$ and $\mathrm{R}$ represent the signals for the left and right ears including HRTFs, but they are described together as $\mathrm{L,R}$ in the subscript to represent both the left and right ear signals, e.g., $h_{\mathrm{L,R}}(\bm{r}_{\mathrm{s}},k)$. 

The HRTF representation in the spherical harmonic domain is generally used for interpolating binaural signals~\cite{god,god2,god3}. We use this representation to obtain a simple formulation of binaural signals on the basis of the harmonic analysis of the soundfield. The HRTFs are measured by loudspeakers at $J$ positions on $\partial\Omega$, whose positions are denoted by $\bm{r}_{\mathrm{s},j}=[R_{\mathrm{s}},\theta_{\mathrm{s},j},\phi_{\mathrm{s},j}]^{\mathsf{T}}$ ($j\in\{1,\ldots,J\}$). We consider to expand $h_{\mathrm{L,R}}(\bm{r}_{\mathrm{s}},k)$ using the complex conjugate of spherical harmonic functions as
\begin{align}
 h_{\mathrm{L,R}}(\bm{r}_{\mathrm{s},j},k) = \sum_{n=0}^{\infty} \sum_{m=-n}^{n} H_{\mathrm{L,R},n}^m(k) Y_n^m(\theta_{\mathrm{s},j},\phi_{\mathrm{s},j})^{\ast}.
\label{eq:hrtf_shd_trun}
\end{align}
By truncating the infinite sum up to $N$, we can describe Eq.~\eqref{eq:hrtf_shd_trun} using the linear equation
\begin{align}
 \bm{h}_{\mathrm{L,R}}(k) = \bm{Y}\bm{\psi}_{\mathrm{L,R}}(k),
\end{align}
where $\bm{h}_{\mathrm{L,R}}(k)\in\mathbb{C}^J$ and $\bm{\psi}_{\mathrm{L,R}}(k)\in\mathbb{C}^{(N+1)^2}$ are the vectors consisting of $h_{\mathrm{L,R}}(\bm{r}_{\mathrm{s},j},k)$ and $H_{\mathrm{L,R},n}^m(k)$, respectively, and $\bm{Y}\in\mathbb{C}^{J \times (N+1)^2}$ is the matrix consisting of spherical harmonic functions $Y_n^m(\theta_{\mathrm{s},j},\phi_{\mathrm{s},j})^{\ast}$. When $J \ge (N+1)^2$, $\bm{\psi}_{\mathrm{L,R}}(k)$ can be obtained as the least-squares solution with Tikhonov regularization as 
\begin{align}
 \bm{\psi}_{\mathrm{L,R}}(k) = \left( \bm{Y}^{\mathsf{H}}\bm{Y} + \gamma \bm{Q} \right)^{-1} \bm{Y}^{\mathsf{H}} \bm{h}_{\mathrm{L,R}}(k),
\label{eq:hrtf_shd_mat}
\end{align}
where $\gamma$ is a constant parameter, $\bm{Q}$ is the weighting diagonal matrix, and the diagonal element for the $n$th-order coefficient of $\bm{Q}$ is $1+n(n+1)$~\cite{god,god3}. 

\section{\label{sec:bnrl_pwd}Problem statement and prior works}

\subsection{\label{subsec:bnrl}Binaural rendering from microphone array signals}

Our objective is to reproduce binaural signals received by a listener at a remote location by using measurements of multiple microphones placed in a recording area. By using the listener's HRTFs (or their proxies), the binaural signals are synthesized so that they are identical to those received when the listener is present in the recording area. The reproduced binaural signals should vary depending on the listener's head position and direction. To reproduce the spatial sound including reverberations, it is necessary to capture the soundfield inside a target listening region by using multiple microphones. Then, the binaural signals are synthesized using HRTFs. 

Suppose that a source-free region of interest $D$ is set in the recording area and $I$ microphones are placed inside $D$. The microphone signals of wavenumber $k$ are denoted as $\bm{s}(k)=[s_1(k), \ldots, s_I(k)]^{\mathsf{T}}$ ($\in \mathbb{C}^I$). The HRTFs from $J$ loudspeakers on a spherical surface $\partial \Omega$ with respect to the listener's left and right eardrums, $h_{\mathrm{L,R}}(\bm{r}_{\mathrm{s}},k)$ ($\bm{r}_{\mathrm{s}} \in \partial \Omega$), are measured beforehand. The goal is to reproduce the binaural signals of the listener, $y_{\mathrm{L,R}}(\bm{r},k)$, when the listener (the origin of the HRTF measurement, i.e., the listener's head center) is positioned at $\bm{r} \in D$. 

This binaural reproduction problem is twofold: soundfield estimation by multiple microphones and binaural rendering using the estimated soundfield. We introduce some prior works in the following sections. Hereafter, the wavenumber $k$ is omitted for notational simplicity.

\subsection{\label{subsec:rgd}Estimation of expansion coefficients using spherical microphone array}

Generally, expansion coefficients of the soundfield with spherical wavefunctions are estimated in the recording area. A single spherical microphone array is typically used for such estimation~\cite{conv2}. To avoid the forbidden frequency problem, microphones mounted on an acoustically rigid baffle, an array of directional microphones, or multiple layers of microphone arrays are usually employed~\cite{Poletti:J_AES_2005,Rafaely:IEEE_J_ASLP_2005,dualsphere,Koyama:JASA_J_2016}. We here assume that the spherical microphone array with a rigid baffle is placed in $D$ with its center at the origin $\bm{r}_{\mathrm{o}}$. 

Since the spherical baffle is assumed to be acoustically rigid, the radial velocity on the surface becomes zero. Then, by denoting the radius of the spherical array as $R_{\mathrm{m}}$, the sound pressure $u(\bm{r})$ ($\bm{r}=[r,\theta,\phi]^{\mathsf{T}} \in D$, $r \geq R_{\mathrm{m}}$) is described by spherical wavefunction expansion as
\begin{align}
 u(\bm{r}) &= \sumsum \alpha_n^m(\bm{r}_{\mathrm{o}}) \notag\\
& \ \ \cdot \sqrt{4\pi} \left( j_n(kr) - \frac{j_n^{\prime}(kR_{\mathrm{m}})}{h^{\prime}_n(kR_{\mathrm{m}})}h_n(kr) \right)Y_n^m(\theta,\phi),
\end{align}
where $\alpha_n^m(\bm{r}_{\mathrm{o}})$ is the interior expansion coefficient at the array center $\bm{r}_{\mathrm{o}}$. Thus, the observation of the $i$th microphone $s_i$ at $\bm{r}_{\mathrm{m},i}=[R_{\mathrm{m}}, \theta_{\mathrm{m},i}, \phi_{\mathrm{m},i}]^{\mathsf{T}}$ is represented as
\begin{align}
 s_i = \sumsum \frac{- \sqrt{4\pi} \mathrm{j}}{k^2 R_{\mathrm{m}}^2 h_n^{\prime}(kR_{\mathrm{m}})} Y_n^m(\theta_{\mathrm{m},i},\phi_{\mathrm{m},i}) \alpha_n^m(\bm{r}_{\mathrm{o}}).
\label{eq:rigsph}
\end{align} 
Here, the following relation is used:
\begin{align}
 j_n(x) h_n^{\prime}(x) - j_n^{\prime}(x) h_n(x) = -\frac{\mathrm{j}}{x^2}.
\end{align} 

One approach to estimating $\alpha_n^m(\bm{r}_{\mathrm{o}})$ up to a predefined truncation order is to derive an analytical representation of $\alpha_n^m(\bm{r}_{\mathrm{o}})$ from Eq.~\eqref{eq:rigsph}~\cite{conv2}, where the integral over the spherical microphone surface must be computed with the discrete observation values $\bm{s}$. We here introduce the least-squares solution of Eq.~\eqref{eq:rigsph} to avoid numerical integration. By truncating the infinite sum up to $N$, Eq.~\eqref{eq:rigsph} can be approximated using the following linear equation:
\begin{align}
\bm{s} = \bm{\Pi} \bar{\bm{\alpha}}(\bm{r}_{\mathrm{o}}),
\end{align}
where $\bar{\bm{\alpha}}(\bm{r}_{\mathrm{o}})\in\mathbb{C}^{(N+1)^2}$ is the vector consisting of $\alpha_n^m(\bm{r}_{\mathrm{o}})$ and $\bm{\Pi}\in\mathbb{C}^{I \times (N+1)^2}$ is the matrix consisting of the coefficients except $\alpha_n^m(\bm{r}_{\mathrm{o}})$ in Eq.~\eqref{eq:rigsph}. When $I \ge (N+1)^2$, $\bar{\bm{\alpha}}(\bm{r}_{\mathrm{o}})$ can be obtained as the least-squares solution with Tikhonov regularization as
\begin{align}
\bar{\bm{\alpha}}(\bm{r}_{\mathrm{o}}) = (\bm{\Pi}^\mathsf{H}\bm{\Pi} + \eta \bm{I})^{-1}\bm{\Pi}^\mathsf{H}\bm{s},
\label{eq:tik_rig}
\end{align} 
where $\eta$ is a constant parameter. This method is applicable only to the estimation of the expansion coefficients at the array center.

\subsection{\label{subsec:bnrl_pwd}Binaural rendering based on plane wave decomposition}

The binaural signals are reproduced from the expansion coefficients estimated in the recording area. Many binaural rendering methods are based on the plane wave decomposition of the soundfield~\cite{conv,conv2,schultz2013data-based}. First, the soundfield $u(\bm{r})$ is represented by the weighted integral of plane waves as
\begin{align}
 u(\bm{r}) = \frac{1}{4\pi} \int_{\mathbb{S}_2} w_{\mathrm{pw}} (\bm{\eta}) e^{\mathrm{j} k \langle \bm{\eta}, \bm{r} \rangle } \mathrm{d} \bm{\eta},
\label{eq:pw_dec}
\end{align}
where $w_{\mathrm{pw}}(\bm{\eta})$ is the weight for the plane wave of the arrival direction $\bm{\eta}=[1, \theta^{\prime}, \phi^{\prime}]^{\mathsf{T}}\in\mathbb{S}_2$. This representation is referred to as the \textit{Herglotz wavefunction}~\cite{singlelayer}. We here reformulate the plane wave expansion using expansion coefficients of the spherical wavefunctions. The plane wave function can be expanded as
\begin{align}
 e^{\mathrm{j} k \langle \bm{\eta}, \bm{r} \rangle} = \sum_{n=0}^{\infty} \sum_{m=-n}^n 4 \pi \mathrm{j}^n j_n(kr) Y_n^m(\theta^{\prime},\phi^{\prime})^{\ast}  Y_n^m(\theta,\phi).
\label{eq:pw_shd}
\end{align}
By substituting Eq.~\eqref{eq:pw_shd} into Eq.~\eqref{eq:pw_dec}, one can obtain
\begin{align}
& u(\bm{r}) = \notag\\
& \ \ \sum_{n=0}^{\infty} \sum_{m=-n}^n \mathrm{j}^n j_n(kr) Y_n^m(\theta,\phi) \int_{\mathbb{S}_2}  w_{\mathrm{pw}}(\bm{\eta}) Y_n^m(\theta^{\prime},\phi^{\prime})^{\ast} \mathrm{d} \bm{\eta}.
\label{eq:pw_dec_shd}
\end{align}
The plane wave weight $w_{\mathrm{pw}}(\bm{\eta})$ can be related to the expansion coefficients $\alpha_n^m(\bm{r})$ in the recording area as
\begin{align}
 w_{\mathrm{pw}}(\bm{\eta}) = \sum_{n=0}^{\infty} \sum_{m=-n}^n \sqrt{4\pi} \mathrm{j}^{-n} \alpha_n^m(\bm{r}) Y_n^m(\theta^{\prime},\phi^{\prime}).
\label{eq:pw_weight_shd}
\end{align}
This relation can be confirmed by substituting Eq.~\eqref{eq:pw_weight_shd} into Eq.~\eqref{eq:pw_dec_shd} and by using the orthogonality of the spherical harmonic functions.

The binaural signals are usually obtained by the summation of $h_{\mathrm{L,R}}(\bm{r}_{\mathrm{s}})$ multiplied by the weight $w_{\mathrm{pw}}(\bm{\eta})$ of each angle. We here obtain a simple formulation by regarding $h_{\mathrm{L,R}}(\bm{r}_{\mathrm{s}})$ as functions of continuous $\bm{r}_{\mathrm{s}} \in \partial\Omega$. We represent the binaural signals at the position $\bm{r}_{\mathrm{o}}$ as 
\begin{align}
 y_{\mathrm{L,R}}(\bm{r}_{\mathrm{o}}) = \frac{1}{R_{\mathrm{s}}^2} \int_{\partial \Omega} w_{\mathrm{pw}}(\hat{\bm{r}}_{\mathrm{s}}) h_{\mathrm{L,R}}(\bm{r}_{\mathrm{s}}) \mathrm{d}\bm{r}_{\mathrm{s}},
\label{eq:bnrl_pw_int}
\end{align}
where $\hat{\bm{r}}_{\mathrm{s}}=\bm{r}_{\mathrm{s}}/\|\bm{r}_{\mathrm{s}}\|$. We expand $h_{\mathrm{L,R}}(\bm{r}_{\mathrm{s}})$ using the complex conjugate of spherical harmonic functions as in Eq.~\eqref{eq:hrtf_shd_trun}. Thus, the binaural signals at $\bm{r}_{\mathrm{o}}$,  $y_{\mathrm{L,R}}(\bm{r}_{\mathrm{o}})$ are represented using $\alpha_n^m(\bm{r}_{\mathrm{o}})$ and $H_{\mathrm{L,R},n}^m$ as
\begin{align}
 y_{\mathrm{L,R}}(\bm{r}_{\mathrm{o}}) = \sum_{n=0}^{\infty} \sum_{m=-n}^n \sqrt{4\pi} \mathrm{j}^{-n} \alpha_n^m(\bm{r}_{\mathrm{o}}) H_{\mathrm{L,R},n}^m.
\label{eq:bnrl_pw}
\end{align}
The infinite sum is truncated up to a predefined truncation order in practice. The distance of the loudspeakers in measuring $h_{\mathrm{L,R}}(\bm{r}_{\mathrm{s}})$, i.e., $R_{\mathrm{s}}$, is assumed to be sufficiently large, and $h_{\mathrm{L,R}}(\bm{r}_{\mathrm{s}})$ is approximated as the transfer function from plane wave sources. 

\section{\label{sec:pro} Proposed method of binaural rendering from distributed microphone array signals}

\subsection{\label{subsec:dismic} Estimation of expansion coefficients using distributed microphone array}

We estimate the expansion coefficients of spherical wavefunctions at the arbitrary position $\bm{r} \in D$ using microphones distributed over $D$. To achieve this, we apply the harmonic analysis of infinite order for the soundfield. The microphone directivities are assumed to be known, and the expansion coefficients of the directivity pattern of the $i$th microphone are denoted by $c_{i,n}^m$. Then, the observation of the $i$th microphone at $\bm{r}_{\mathrm{m},i}$, $s_i$, is described using the expansion coefficients of the soundfield as
\begin{align}
s_i &= \sum_{n=0}^{\infty} \sum_{m=-n}^n c_{i,n}^{m \ast} \alpha_n^m(\bm{r}_{\mathrm{m},i}) \notag\\
&=\bm{c}_i^\mathsf{H}\bm{\alpha}(\bm{r}_{\mathrm{m},i}) \notag\\
&=\bm{c}_i^\mathsf{H} \bm{T}(\bm{r}_{\mathrm{m},i} - \bm{r}) \bm{\alpha}(\bm{r}),
\label{eq:mic_m}
\end{align}
where $\bm{\alpha}(\bm{r})\in\mathbb{C}^{\infty}$ and $\bm{c}_i\in\mathbb{C}^{\infty}$ are the infinite-dimensional vectors consisting of $\alpha_n^m(\bm{r})$ and $c_{i,n}^m$, respectively. Equation~\eqref{eq:mic_m} can be rewritten as
 \begin{align}
\bm{s} = \bm{\Xi}(\bm{r})^\mathsf{H} \bm{\alpha}(\bm{r}),
\end{align}
where $\bm{\Xi}(\bm{r})\in\mathbb{C}^{\infty \times I}$ is obtained as 
\begin{align}
\bm{\Xi}(\bm{r}) &= [(\bm{c}_1^\mathsf{H}\bm{T}(\bm{r}_{\mathrm{m},1}-\bm{r}))^\mathsf{H} \cdots (\bm{c}_I^\mathsf{H}\bm{T}(\bm{r}_{\mathrm{m},I}-\bm{r}))^\mathsf{H}] \nonumber \\
&=[\bm{T}(\bm{r}-\bm{r}_{\mathrm{m},1})\bm{c}_1 \cdots \bm{T}(\bm{r}-\bm{r}_{\mathrm{m},I})\bm{c}_I].
\label{eq:xi}
\end{align}
We used the property of the translation operator \eqref{eq:trans_prop} to derive Eq.~\eqref{eq:xi}. The expansion coefficients $\bm{\alpha}(\bm{r})$ is estimated as~\cite{Ueno:IEEE_SPL2018}
\begin{align}
\bm{\alpha}(\bm{r}) = \bm{\Xi}(\bm{r})(\bm{\Psi} + \lambda \bm{I})^{-1} \bm{s},
\label{eq:alpha_lam}
\end{align}
where $\lambda$ is a constant parameter and $\bm{\Psi}=\bm{\Xi}(\bm{r})^\mathsf{H} \bm{\Xi}(\bm{r}) \in \mathbb{C}^{I \times I}$. From the property in Eq.~\eqref{eq:trans_prop}, the $(i,i^{\prime})$th element of $\bm{\Psi}$ becomes
\begin{align}
 (\bm{\Psi})_{i,i^{\prime}} &= \bm{c}_i^{\mathsf{H}} \bm{T}(\bm{r}_{\mathrm{m},i}-\bm{r}) \bm{T}(\bm{r}-\bm{r}_{\mathrm{m},i^{\prime}}) \bm{c}_{i^{\prime}} \notag\\
&= \bm{c}_i^{\mathsf{H}} \bm{T}(\bm{r}_{\mathrm{m},i} - \bm{r}_{\mathrm{m},i^{\prime}}) \bm{c}_{i^{\prime}}. 
\end{align}
Therefore, $\bm{\Psi}$ does not depend on the position $\bm{r}$ and depends only on the microphone positions and directivities. Note that $\bm{c}_i$ is typically modeled by low-order coefficients. 

\subsection{\label{subsec:bnrl_sph}Binaural rendering based on spherical wave decomposition}

To consider the loudspeaker distance in the HRTF measurement, we represent the binaural signals assuming that $h_{\mathrm{L,R}}(\bm{r}_{\mathrm{s}})$ is the transfer function from the point source. Thus, the soundfield $u(\bm{r})$ is represented as the weighted integral of spherical waves from a point source on $\partial \Omega$ as
\begin{align}
 u(\bm{r}) = \int_{\partial\Omega} w_{\mathrm{sph}}(\bm{r}_{\mathrm{s}}) G(\bm{r}-\bm{r}_{\mathrm{s}}) \mathrm{d} \bm{r}_{\mathrm{s}},
\label{eq:slp}
\end{align}
where $G(\bm{r}-\bm{r}_{\mathrm{s}})$ is the transfer function of the point source, which is equivalent to the three dimensional (3D) free-field Green's function defined as~\cite{acous}
\begin{align}
& G(\bm{r}-\bm{r}_{\mathrm{s}}) = \frac{e^{-\mathrm{j}k \|\bm{r}-\bm{r}_{\mathrm{s}}\|}}{4\pi \|\bm{r}-\bm{r}_{\mathrm{s}}\|} \notag\\
& \ \ = \sum_{n=0}^{\infty} \sum_{m=-n}^{n} -\mathrm{j} k j _n(kr) h_n(kR_{\mathrm{s}}) Y^m_n(\theta,\phi) Y_n^{m}(\theta_{\mathrm{s}},\phi_{\mathrm{s}})^{\ast}.
\label{eq:green}
\end{align}
We here use the same spherical surface $\partial \Omega$ as in the HRTF measurements, and $\bm{r}_{\mathrm{s}}$ is the position vector on $\partial \Omega$. The representation in Eq.~\eqref{eq:slp} is referred to as the \textit{single-layer potential}~\cite{singlelayer} or \textit{simple source formulation}\cite{acous}. The spherical wave weight $w_{\mathrm{sph}}(\bm{r}_{\mathrm{s}})$ can be related to the expansion coefficients $\alpha_n^m(\bm{r})$ as
\begin{align}
 w_{\mathrm{sph}}(\bm{r}_{\mathrm{s}}) = \sum_{n=0}^{\infty} \sum_{m=-n}^n \frac{\sqrt{4\pi} \mathrm{j}}{kR_{\mathrm{s}}^2 h_n(kR_{\mathrm{s}})} \alpha_n^m(\bm{r}) Y_n^m(\theta_{\mathrm{s}},\phi_{\mathrm{s}}).
\label{eq:sph_weight_shd}
\end{align}
This relation can be confirmed by substituting Eqs.~\eqref{eq:green} and \eqref{eq:sph_weight_shd} into Eq.~\eqref{eq:slp} and using the orthogonality of the spherical harmonic function.

Finally, the binaural signals at $\bm{r}$, $y_{\mathrm{L,R}}(\bm{r})$, are obtained using $\alpha_n^m(\bm{r})$ and $H_{\mathrm{L,R},n}^m$ as
\begin{align}
 y_{\mathrm{L,R}}(\bm{r}) &= \int_{\partial\Omega} w_{\mathrm{sph}}(\bm{r}_{\mathrm{s}}) h_{\mathrm{L,R}}(\bm{r}_{\mathrm{s}}) \mathrm{d} \bm{r}_{\mathrm{s}} \notag\\
&= \sum_{n=0}^{\infty} \sum_{m=-n}^n \frac{\sqrt{4\pi} \mathrm{j}}{k h_n(kR_{\mathrm{s}})} \alpha_n^m(\bm{r}) H_{\mathrm{L,R},n}^m.
\label{eq:bnrl_sph}
\end{align}
The infinite sum is truncated up to a finite order in practice. It can be confirmed that Eq.~\eqref{eq:bnrl_sph} corresponds to Eq.~\eqref{eq:bnrl_pw} in the plane-wave-decomposition-based method when $R_{\mathrm{s}}$ becomes infinity because $h_n(kR_{\mathrm{s}})$ can be approximated as $\mathrm{j}^{n+1}/k$ for far-field directivity ($R_{\mathrm{s}} \to \infty$)~\cite{acous}. Similar formulations have been derived to take into consideration the loudspeaker distance in the context of near-field compensation of higher order Ambisonics~\cite{Daniel:AES116conv} and HRTF interpolation~\cite{god,Pollow:Acustica2012}. In Sect.~\ref{sec:pln_vs_sph}, it is shown that this distance-compensated rendering method significantly improves the amplitude response of the reproduced binaural signals.

\subsection{\label{subsec:trans}Adaptation to rotation and translation of listener's head}

The binaural signals should be reproduced with adaptation to the listener's head movement. In the plane-wave-decomposition-based methods, the binaural signals are adapted to the listening position by the phase shift of plane wave weights~\cite{conv,plane_trans}. When the expansion center is moved from the origin to $\bm{r}^{\prime}$, Eq.~\eqref{eq:pw_dec} is transformed to
\begin{align}
 u(\bm{r}) = \frac{1}{4\pi}\int_{\mathbb{S}_2} w_{\mathrm{pw}}(\bm{\eta}) e^{\mathrm{j} k \langle \bm{\eta}, \bm{r}^{\prime} \rangle}  e^{\mathrm{j} k \langle \bm{\eta}, \bm{r}-\bm{r}^{\prime} \rangle} \mathrm{d}\bm{\eta}.
\end{align}
Thus, the plane wave weight at the shifted position $\bm{r}^{\prime}$ is obtained as
\begin{align}
 w_{\mathrm{pw}}(\bm{\eta};\bm{r}^{\prime}) = w_{\mathrm{pw}}(\bm{\eta})e^{\mathrm{j} k \langle \bm{\eta}, \bm{r}^{\prime} \rangle }.
\end{align}
To rotate the binaural signals, Eq.~\eqref{eq:bnrl_pw_int} is computed using HRTFs for rotated angles. When the HRTFs of the rotated angles are not available, some interpolation techniques are necessary~\cite{god,god2,god3}. 

In the methods based on the spherical wavefunction expansion, the translation and rotation of the listening position are achieved by transforming the estimated expansion coefficients. The center of the expansion coefficients is changed from $\bm{r}$ to $\bm{r}^{\prime}$ using the translation operator in Eq.~\eqref{eq:trans_coef}. In the truncation-based methods as described in Sect.~\ref{subsec:rgd}, this translation operator is calculated by truncating $\bm{T}(\cdot)$ up to a predefined order. On the other hand, in the proposed method, the translation is calculated as
  \begin{align}
 \bm{\alpha}(\bm{r}^{\prime}) &= \bm{T}(\bm{r}^{\prime}-\bm{r}) \bm{\alpha}(\bm{r})  \notag\\
&= \bm{T}(\bm{r}^{\prime}-\bm{r}) \bm{\Xi}(\bm{r})(\bm{\Psi} + \lambda \bm{I})^{-1} \bm{s} \notag\\
&= \bm{\Xi}(\bm{r}^{\prime})(\bm{\Psi} + \lambda \bm{I})^{-1} \bm{s}.
\end{align}
Since $\bm{\Psi}$ does not depend on the position $\bm{r}$ or $\bm{r}^{\prime}$, the expansion coefficients at the listening position can be obtained by changing $\bm{\Xi}(\bm{r})$ to $\bm{\Xi}(\bm{r}^{\prime})$ with fixed $(\bm{\Psi} + \lambda \bm{I})^{-1}\bm{s}$. Thus, the spherical wave weight at the shifted position $\bm{r}^{\prime}$, $w_{\mathrm{sph}}(\bm{r}_{\mathrm{s}};\bm{r}^{\prime})$, is obtained by replacing $\bm{\alpha}(\bm{r})$ in Eq.~\eqref{eq:bnrl_sph} with $\bm{\alpha}(\bm{r}^{\prime})$.

Adaptation to the head rotation is achieved using the rotated expansion coefficients described with the Euler angles $(\mathring{\alpha},\mathring{\beta},\mathring{\gamma})$ as
\begin{align}
 \alpha_n^{m^{\prime}}(\bm{r}; \mathring{\alpha}, \mathring{\beta}, \mathring{\gamma}) = \sum_{m=-n}^{n} D_{m^{\prime},m}^n(\mathring{\alpha},\mathring{\beta},\mathring{\gamma}) \alpha_n^m(\bm{r}),
\label{eq:rot_coef}
\end{align}
where $D_{m,m^{\prime}}^n(\cdot)$ is the element of the Wigner-$D$ matrix~\cite{Edmonds}. By using the Wigner $D$-matrix $\bm{R}(\mathring{\alpha},\mathring{\beta},\mathring{\gamma})\in\mathbb{C}^{\infty \times \infty}$ consisting of $D_{m,m^{\prime}}^n(\cdot)$, we can represent Eq.~\eqref{eq:rot_coef} in a matrix form as
\begin{align}
 \bm{\alpha}(\bm{r}; \mathring{\alpha}, \mathring{\beta}, \mathring{\gamma}) = \bm{R}(\mathring{\alpha}, \mathring{\beta}, \mathring{\gamma}) \bm{\alpha}(\bm{r}). 
\end{align}
The expansion coefficient $\bm{\alpha}(\bm{r})$ is estimated up to a finite order in practice. When the truncation order of $\bm{\alpha}(\bm{r})$ is $N$, $\bm{R}(\mathring{\alpha},\mathring{\beta},\mathring{\gamma})$ becomes a $(N+1)^2 \times (N+1)^2$ block diagonal matrix~\cite{Edmonds}.

Our proposed method is summarized as follows. When the estimated expansion coefficients are truncated up to $N$, the binaural signals at the listening position $\bm{r}$ with the rotation angles $(\mathring{\alpha},\mathring{\beta},\mathring{\gamma})$ are represented as
\begin{align}
y_\mathrm{L,R}(\bm{r}; \mathring{\alpha}, \mathring{\beta}, \mathring{\gamma}) = \bm{\psi}_{\mathrm{L,R}}^\mathsf{T} \bm{\Lambda} \bar{\bm{R}}(\mathring{\alpha},\mathring{\beta},\mathring{\gamma}) \bar{\bm{\Xi}}(\bm{r})(\bm{\Psi} + \lambda \bm{I})^{-1} \bm{s},
\label{eq:pro_method}
\end{align}
where $\bar{\bm{\Xi}} \in \mathbb{C}^{(N+1)^2 \times I}$ and $\bar{\bm{R}} \in \mathbb{C}^{(N+1)^2 \times (N+1)^{2}}$ are constructed by truncating $\bm{\Xi}$ and $\bm{R}$, respectively, and $\bm{\Lambda}\in\mathbb{C}^{(N+1)^2 \times (N+1)^2}$ is a diagonal matrix consisting of $\sqrt{4\pi}\mathrm{j}/kh_n(kR_{\mathrm{s}})$. Since the conversion from microphone observations to binaural signals is a linear operation, $y_{\mathrm{L,R}}$ is obtained by applying the multiple--input--multiple--output (MIMO) finite impulse response (FIR) filter to $\bm{s}$ in practical implementations. 

\section{\label{sec:pln_vs_sph} Comparison of binaural rendering methods}

We first compare binaural rendering accuracy between plane-wave-decomposition-based and spherical-wave-decomposition-based methods described in Sects.~\ref{subsec:bnrl_pwd} and \ref{subsec:bnrl_sph}, which are hereafter denoted as \textbf{PLN} and \textbf{SPH}, respectively. The expansion coefficient of the soundfield in the recording area $\alpha_n^m(\bm{r})$ is assumed to be given in this section. 

HRTFs were obtained by numerical computation using the boundary element method (BEM). The software Mesh2HRTF~\cite{mesh2hrtf,mesh} was used. The head shape of a subject was scanned by magnetic resonance imaging (MRI), and the mesh data of $25968$ surface elements were used to generate HRTFs. On the basis of reciprocity, HRTFs were obtained as transfer functions from the point source at the entrance of the ear canal to the sampling points on the spherical surface $\partial\Omega$. An acoustically rigid surface of the head was assumed. The radius of $\partial \Omega$, $R_{\mathrm{s}}$, was $1.5~\mathrm{m}$, which is the typically used distance for HRTF measurements~\cite{datahrtf,datahrtf2}. The sampling points on $\partial\Omega$ were from $0^\circ$ to $355^\circ$ for the azimuth angle and from $10^\circ$ to $160^\circ$ for the zenith angle at intervals of $5^{\circ}$; therefore, the total number of sampling points was $2232$. The sound velocity was $346.2~\mathrm{m/s}$. The frequency range of interest was set as $100$--$15000~\mathrm{Hz}$ at intervals of $100~\mathrm{Hz}$. 

We assumed that a single point source was positioned at $\bm{r}_{\mathrm{ps}}=[r_{\mathrm{ps}},\theta_{\mathrm{ps}},\phi_{\mathrm{ps}}]^{\mathsf{T}}$ in the free-field recording area, and the expansion coefficient $\alpha_n^m(\bm{r})$ was analytically calculated. Thus, true binaural signals were also obtained as the HRTF from the point source at $\bm{r}_{\mathrm{ps}}$ using BEM. The truncation order for HRTFs in Eq.~\eqref{eq:hrtf_shd_trun} was set as $35$. The truncation order $N$ for the binaural rendering in Eqs.~\eqref{eq:bnrl_pw} and \eqref{eq:bnrl_sph} was set as $N=\min(\lceil ekR_{\mathrm{w}}/2 \rceil, 35)$~\cite{Ward:IEEE_J_SAP2001}, where $e$ is Napier's constant and $R_{\mathrm{w}} = 0.45~\mathrm{m}$ is the average shoulder width of Japanese males.

Binaural signals were evaluated for the left ear from the point sources on the horizontal plane ($\theta_{\mathrm{ps}}=90^\circ$). We here added the additional arguments $r_{\mathrm{ps}}$ and $\phi_{\mathrm{ps}}$ to $y_{\mathrm{L}}(k)$ to indicate the location of the virtual point source of each binaural signal. For evaluation, we introduce two measures. One is the normalized mean square error (NMSE) defined as
\begin{align}
&\mathrm{NMSE}(r_{\mathrm{ps}},\phi_{\mathrm{ps}},k_f) =  \notag \\
& \hspace{20pt} 10 \log_{10} \frac{|{y}_{\mathrm{L}}^{\mathrm{est}}(k_f; r_{\mathrm{ps}},\phi_{\mathrm{ps}}) - {y}_{\mathrm{L}}^{\mathrm{true}}(k_f; r_{\mathrm{ps}},\phi_{\mathrm{ps}})|^2}{|{y}_{\mathrm{L}}^{\mathrm{true}}(k_f; r_{\mathrm{ps}},\phi_{\mathrm{ps}})|^2},
\label{eq:NMSE}
\end{align}
where ${y}_{\mathrm{L}}^{\mathrm{true}}$ and ${y}_{\mathrm{L}}^{\mathrm{est}}$ are the true and estimated binaural signals, respectively, and the subscript $f$ denotes the index of frequency bins. The other is the spectral distortion (SD) defined as~\cite{SD}
\begin{equation}
\mathrm{SD}(r_{\mathrm{ps}},\phi_{\mathrm{ps}}) =  \sqrt{\frac{1}{F}\sum_{f} \left( 20 \log_{10} \frac{|{y}_{\mathrm{L}}^{\mathrm{est}}(k_f; r_{\mathrm{ps}},\phi_{\mathrm{ps}})|}{|{y}_{\mathrm{L}}^{\mathrm{true}}(k_f; r_{\mathrm{ps}},\phi_{\mathrm{ps}})|} \right)^2},
\end{equation}
where only the difference in amplitude response is evaluated. To compare the methods based on plane wave and spherical wave decomposition, the amplitudes of the binaural signals generated by \textbf{PLN} and \textbf{SPH} were normalized to remove the amplitude bias as
\begin{align}
 \hat{y}_{\mathrm{L}}(k_f; r_{\mathrm{ps}},\phi_{\mathrm{ps}}) = \frac{y_{\mathrm{L}}(k_f; r_{\mathrm{ps}},\phi_{\mathrm{ps}})}{\sqrt{\frac{1}{F} \sum_f |y_{\mathrm{L}}(k_f; r_{\mathrm{ps}},\phi_{\mathrm{ps}})|^2 }},
\end{align}
where $F$ is the number of frequency bins. The phase difference was also corrected in the time domain using the cross correlation between the true and reproduced signals. 

\begin{figure}[t]
\figcolumn{
\fig{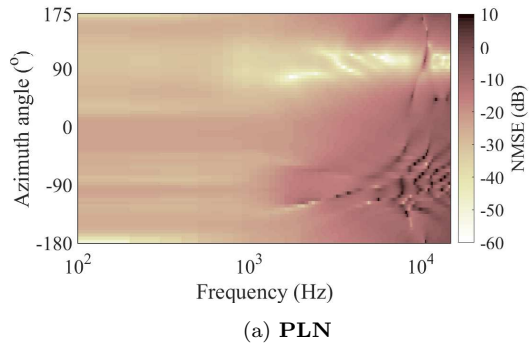}{.4\textwidth}{(a) \textbf{PLN}}
\fig{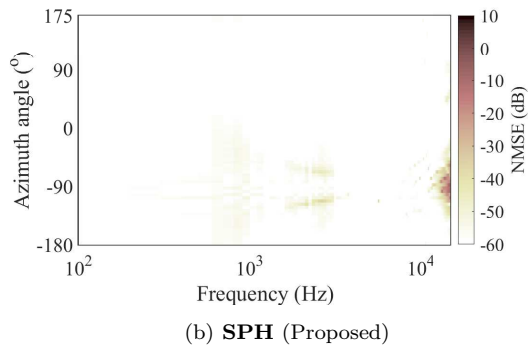}{.4\textwidth}{(b) \textbf{SPH} (Proposed)}
}
\caption{NMSEs with respect to azimuth angle of point source at $r_{\mathrm{ps}}=2.0~\mathrm{m}$ and frequency.}
\label{fig:nmse}
\end{figure}

\begin{figure}[t]
\begin{center}
\includegraphics[width=7.0cm]{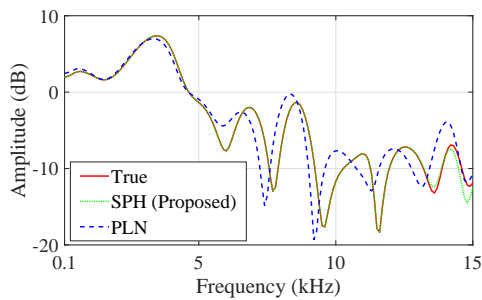}
\end{center}
\caption{Amplitude response of binaural signal of left ear for point source at $(2.0~\mathrm{m},90^{\circ},-90^{\circ})$ with respect to frequency.}
\label{spec}
\end{figure}

Figure~\ref{fig:nmse} shows the NMSEs of the binaural signals generated by \textbf{PLN} and \textbf{SPH} with respect to the azimuth angle of the point source at $r_{\mathrm{ps}}=2.0~\mathrm{m}$ and frequency. A generally lower NMSE was achieved by \textbf{SPH} than by \textbf{PLN}. In particular, the NMSE of \textbf{PLN} was relatively large at high frequencies. As an example, the amplitude response of the binaural signal for the point source at $(2.0~\mathrm{m},90^{\circ},-90^{\circ})$ with respect to frequency is plotted in Fig.~\ref{spec}. The dips in the amplitude response were accurately reproduced by \textbf{SPH}. On the other hand, the frequencies having dips for \textbf{PLN} were shifted from those for the true response at several frequencies, e.g., 7.7~kHz. 

\begin{figure}[t]
\begin{center}
\includegraphics[width=7.0cm]{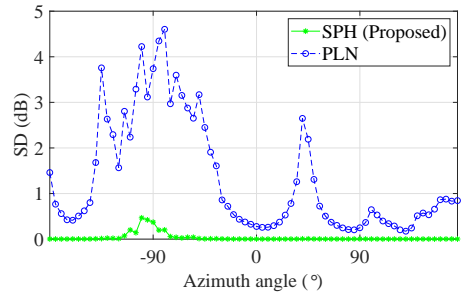}
\caption{SD with respect to azimuth angle of point source at $r_{\mathrm{ps}}=2.0~\mathrm{m}$.}
\label{fig:SD}
\end{center}
\end{figure}
\begin{figure}[t]
\begin{center}
\includegraphics[width=7.0cm]{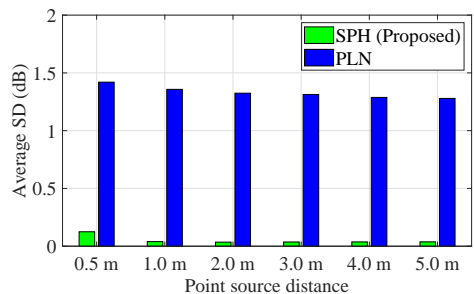}
\caption{SD with respect to point source distance.}
\label{fig:SD_dist}
\end{center}
\end{figure}

The SD is plotted for $r_{\mathrm{ps}}=2.0~\mathrm{m}$ in Fig.~\ref{fig:SD} to evaluate general accuracy with respect to the azimuth angle. Again, a low SD was achieved by \textbf{SPH}. The SD of \textbf{PLN} was particularly high at around $\phi_{\mathrm{ps}}=-90^{\circ}$, which is the point source direction of the opposite side of the head from the left ear where the direct sound is relatively weak. The SD averaged for the azimuth angle with respect to the source distance $r_{\mathrm{ps}}$ is shown in Fig.~\ref{fig:SD_dist}. The average SD of \textbf{SPH} was lower than that of \textbf{PLN} for all distances. 

\section{\label{sec:distmic}Design and evaluation of composite microphone array}

\begin{figure*}[!t]
\begin{center}
\figline{
\fig{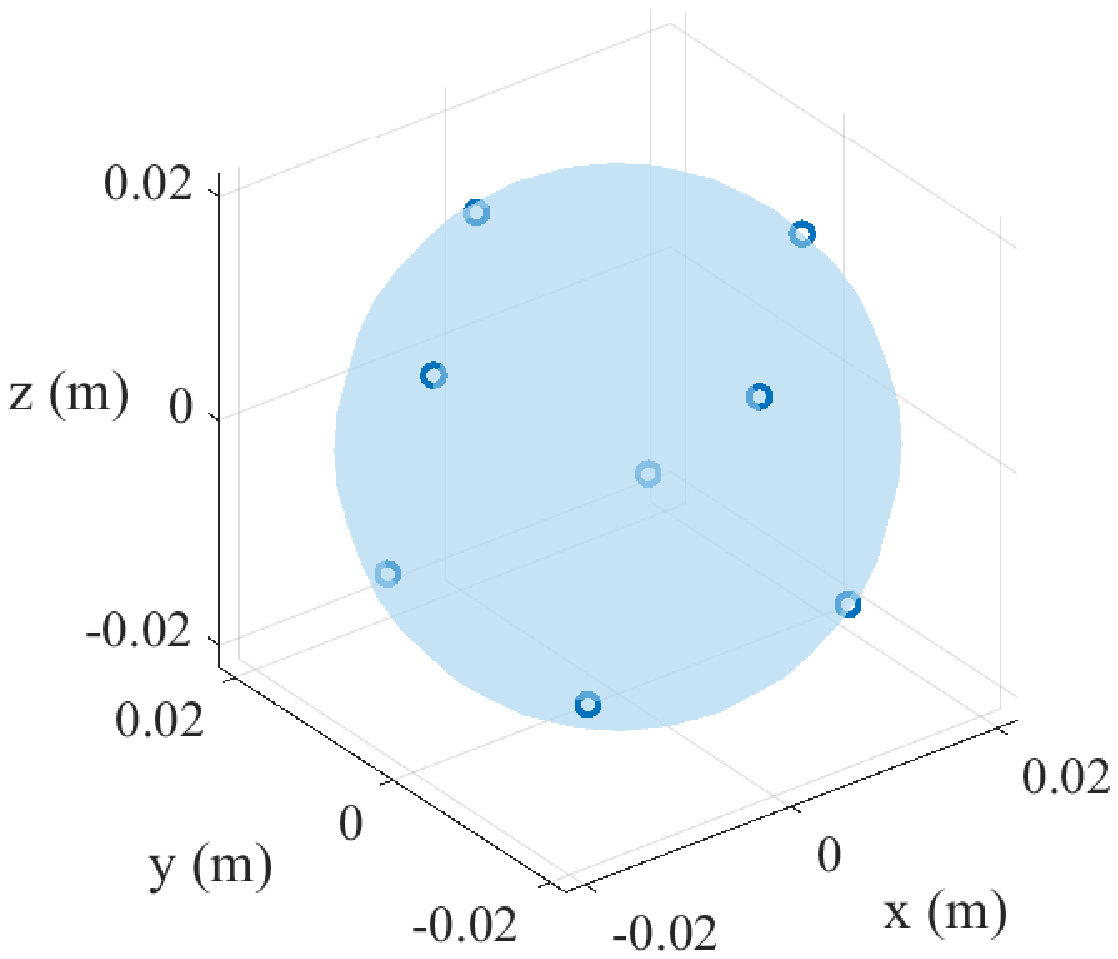}{2.2in}{(a) Single array}
\fig{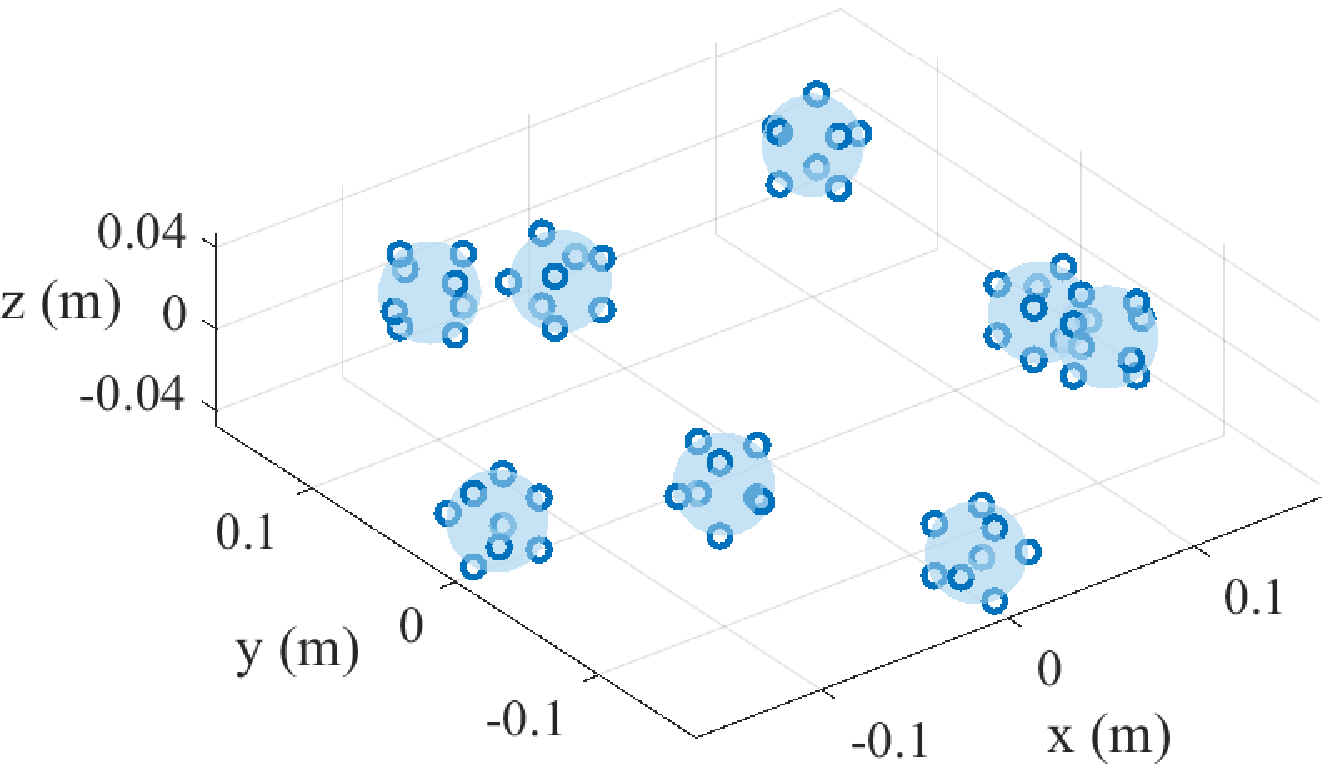}{2.2in}{(b) Composite array}
\fig{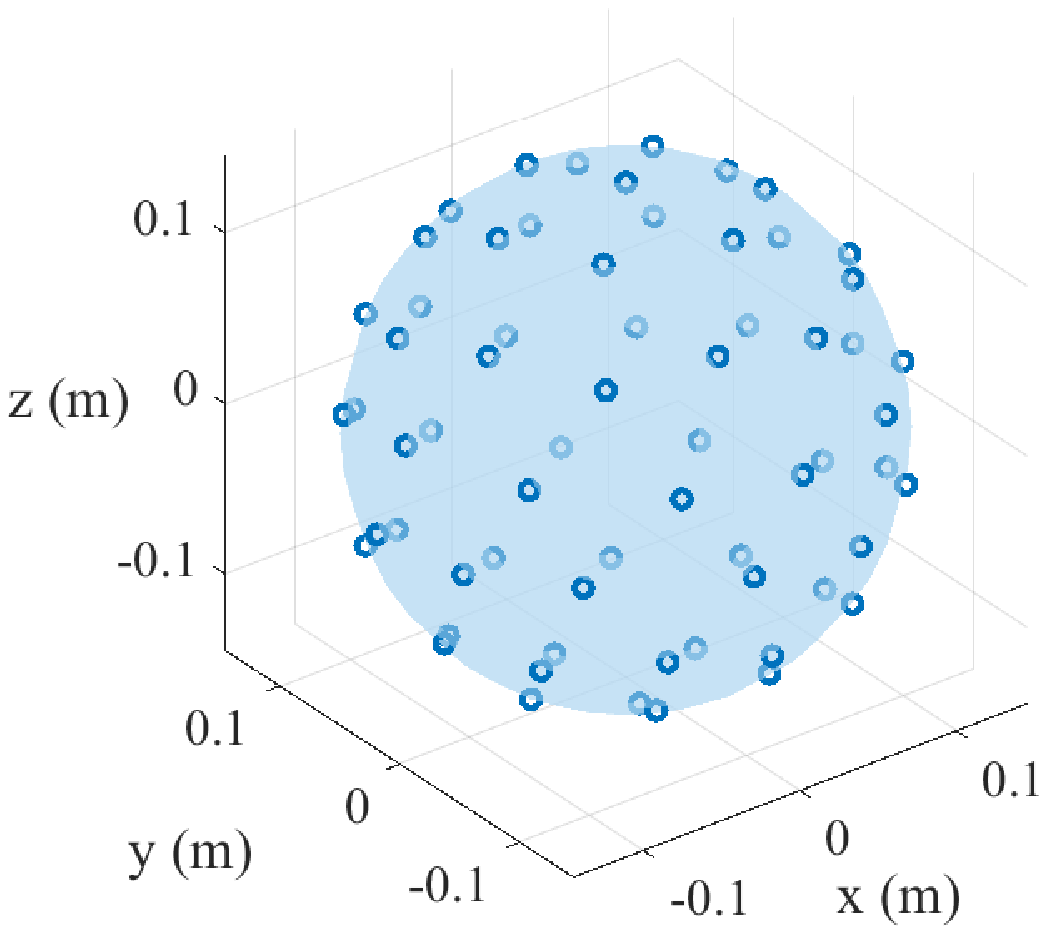}{2.2in}{(c) Spherical array}
}
\caption{Geometry of microphone arrays for (a) single, (b) composite, and (c) spherical arrays.}
\label{fig:micarray}
\end{center}
\end{figure*}

Although the array of microphones (nearly) uniformly arranged on a single sphere can efficiently capture the soundfield, particularly for estimating the expansion coefficients of the spherical wavefunctions at the array center, it does not have flexibility in the microphone arrangement. One of the benefits of our proposed soundfield estimation method based on infinite dimensional analysis is its applicability to a wide range of microphone array geometries. Therefore, we consider using multiple small microphone arrays. Such a composite array system will have flexibility in the array placement and scalability for adding/removing small microphone arrays, depending on the setting of the region of interest $D$. We evaluate our soundfield estimation method using a composite microphone array by numerical simulations. 

For each small microphone array, we used one array of eight unidirectional microphones as shown in Fig.~\ref{fig:micarray}(a), whose arrangement was identical to that of commercially available 2nd-order ambisonics microphones (the shape of tetragonal trapezohedron)~\cite{OctoMic,ambisonic_3Dprint}. The composite microphone array consisted of eight small microphone arrays as shown in Fig.~\ref{fig:micarray}(b), so the total number of microphones was 64. To cover the listening area, which should be larger than the average size of the listener's head, four small arrays were equiangularly placed on the circle of $0.145~\mathrm{m}$ radius at two heights $z=\pm 0.025~\mathrm{m}$. For comparison, we also employed a spherical array of omnidirectional microphones mounted on a rigid baffle of $0.145~\mathrm{m}$ radius, where the arrangement of 64 microphones was determined by the spherical $t$-design for $t=7$~\cite{t-design} (Fig.~\ref{fig:micarray}(c)). 

The directivity pattern of the $i$th unidirectional microphone is represented as
\begin{align}
 c_i(\theta_{\mathrm{m},i},\phi_{\mathrm{m},i}) = \beta + (1-\beta) \langle \bm{\eta},\bm{\eta}_{\mathrm{m},i}\rangle, 
\end{align}
where $\beta \in [0,1]$ is a constant parameter, $\bm{\eta}_{\mathrm{m},i}$ and $\bm{\eta}$ are the directions of the $i$th unidirectional microphone (peak of directivity) and incident soundwave, respectively. Thus, the expansion coefficients of the directivity pattern are derived as
\begin{align}
\bm{c}_i 
&=[c_{i,0}^0,c_{i,1}^{-1},c_{i,1}^0,c_{i,1}^1]^\mathsf{T} \nonumber \\
&=\left[\beta, \frac{\sqrt{4\pi}\mathrm{j}}{3}(1-\beta)Y_1^{-1}(\theta_{\mathrm{m},i},\phi_{\mathrm{m},i})^*, \right. \notag\\
& \hspace{0.5cm} \frac{\sqrt{4\pi}\mathrm{j}}{3} (1-\beta)Y_1^{0}(\theta_{\mathrm{m},i},\phi_{\mathrm{m},i})^{\ast}, \notag\\
& \hspace{0.5cm} \left. \frac{\sqrt{4\pi}\mathrm{j}}{3}(1-\beta)Y_1^{1}(\theta_{\mathrm{m},i},\phi_{\mathrm{m},i})^{\ast} \right]^\mathsf{T}.
\label{eq:uni_dir_coef}
\end{align}
We set $\beta=1/2$ in the experiments described in this section.

We evaluated the binaural reproduction accuracy using the composite microphone array (Fig.~\ref{fig:micarray}(b)) and spherical microphone array (Fig.~\ref{fig:micarray}(c)). Our proposed soundfield estimation method based on infinite-dimensional analysis described in Sect.~\ref{subsec:dismic} was used for the composite array. For the spherical array, the truncation-based estimation method in Sect.~\ref{subsec:rgd} was applied. The binaural rendering method from the estimated expansion coefficients was the \textbf{SPH} for both arrays. The HRTF dataset of NEUMANN KU-100~\cite{lebedev_hrtf} was used, which was measured for $2702$ loudspeaker positions on a sphere of $1.5~\mathrm{m}$ radius. The truncation order for the binaural rendering in Eq.~\eqref{eq:hrtf_shd_trun} was set as $7$ since the number of microphones was 64.

\begin{figure}[t!]
\begin{center}
 \figline{\leftfig{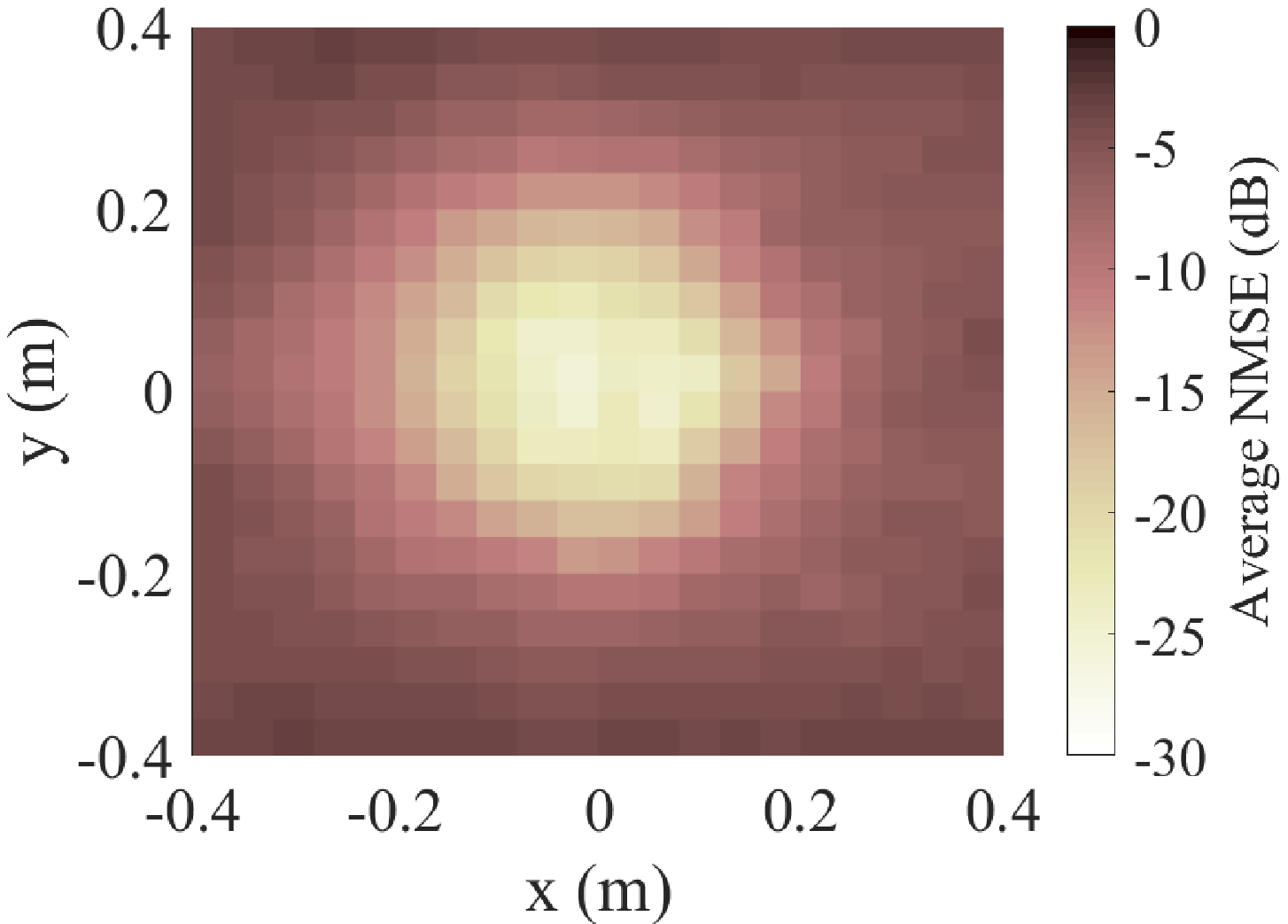}{.25\textwidth}{(a) Composite array}
\leftfig{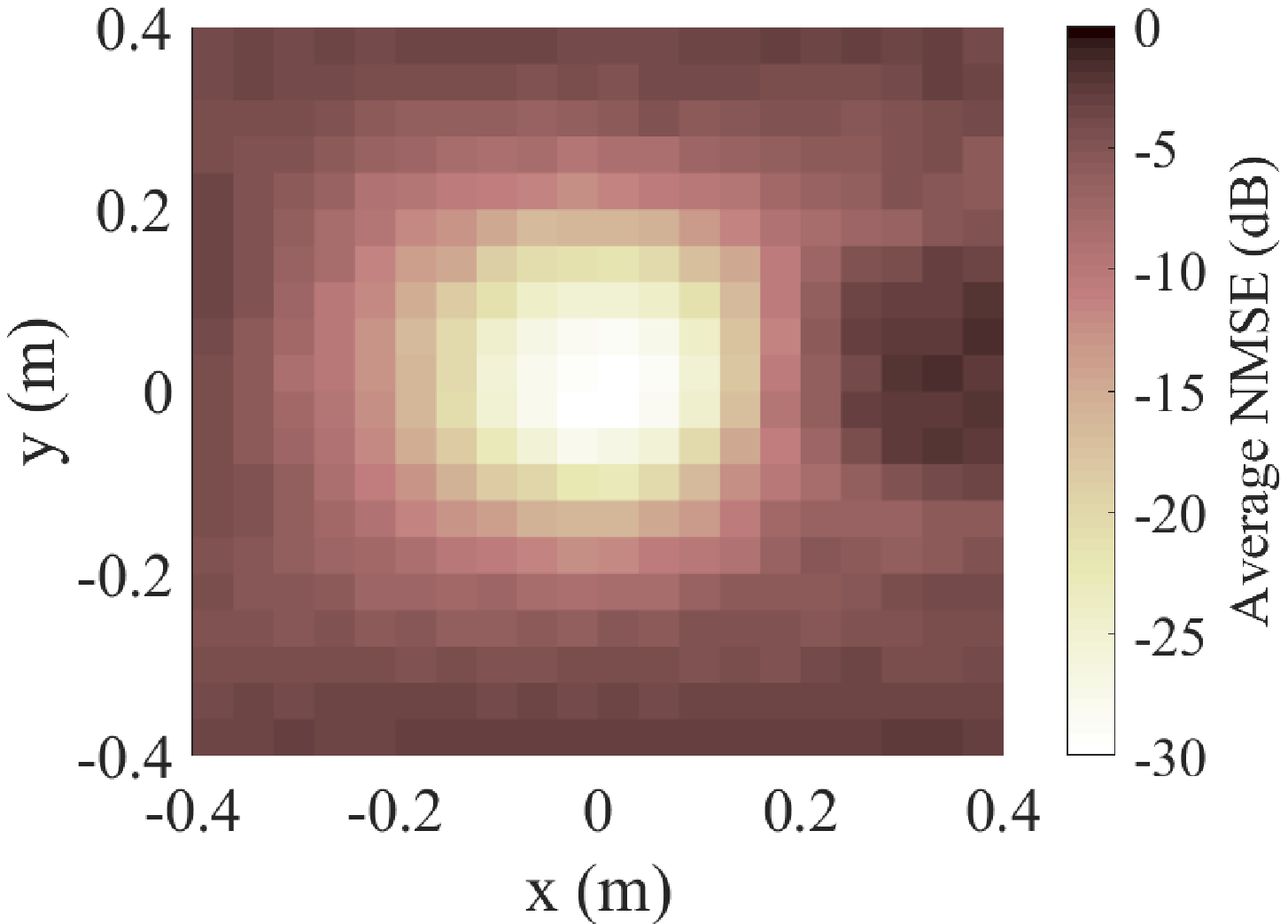}{.25\textwidth}{(b) Spherical array}\hfill}
\end{center}
\caption{Distribution of average NMSE on $x$-$y$-plane at $z=0$.}
\label{fig:ave_NMSE_xy}
\begin{center}
\figline{\leftfig{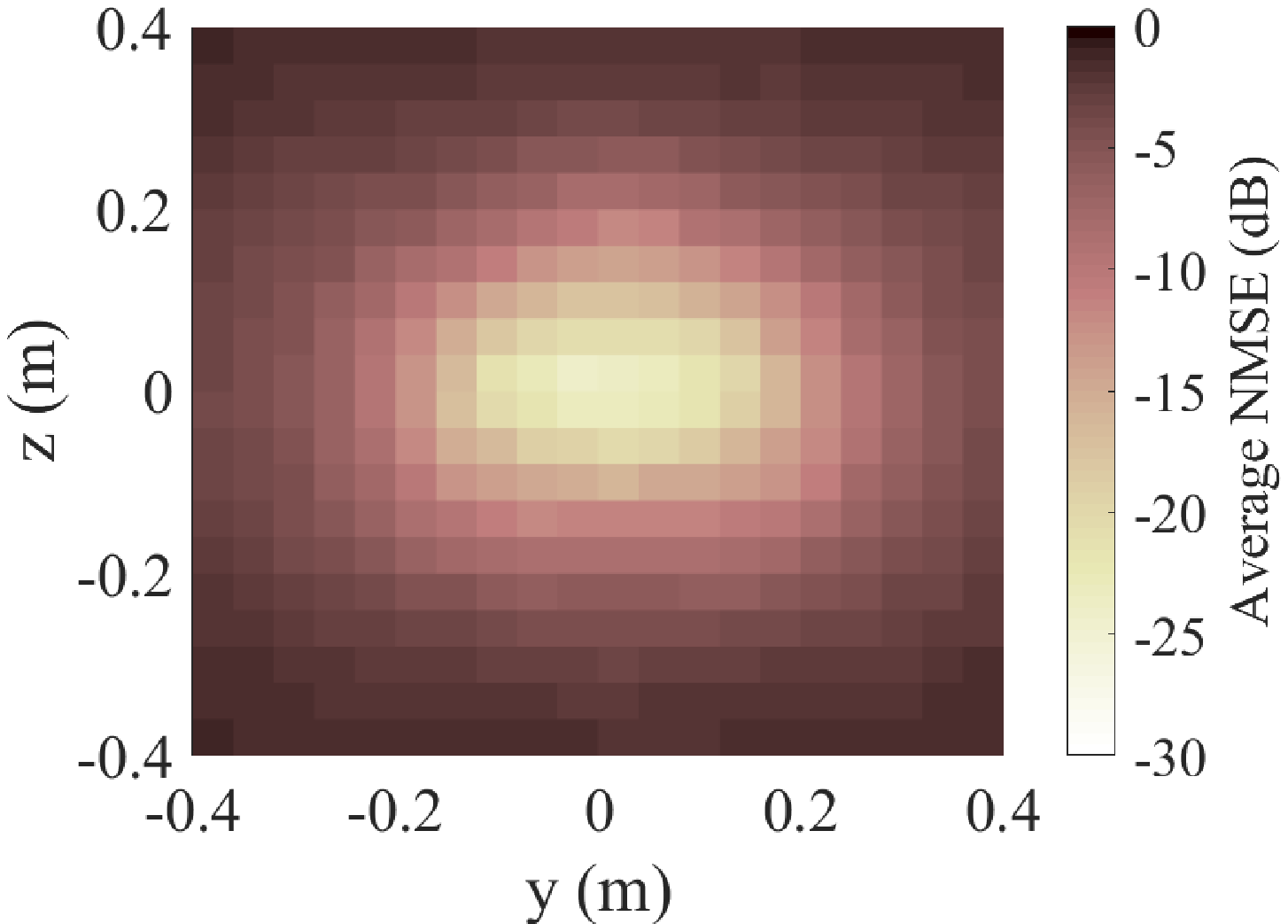}{.25\textwidth}{(a) Composite array}
\leftfig{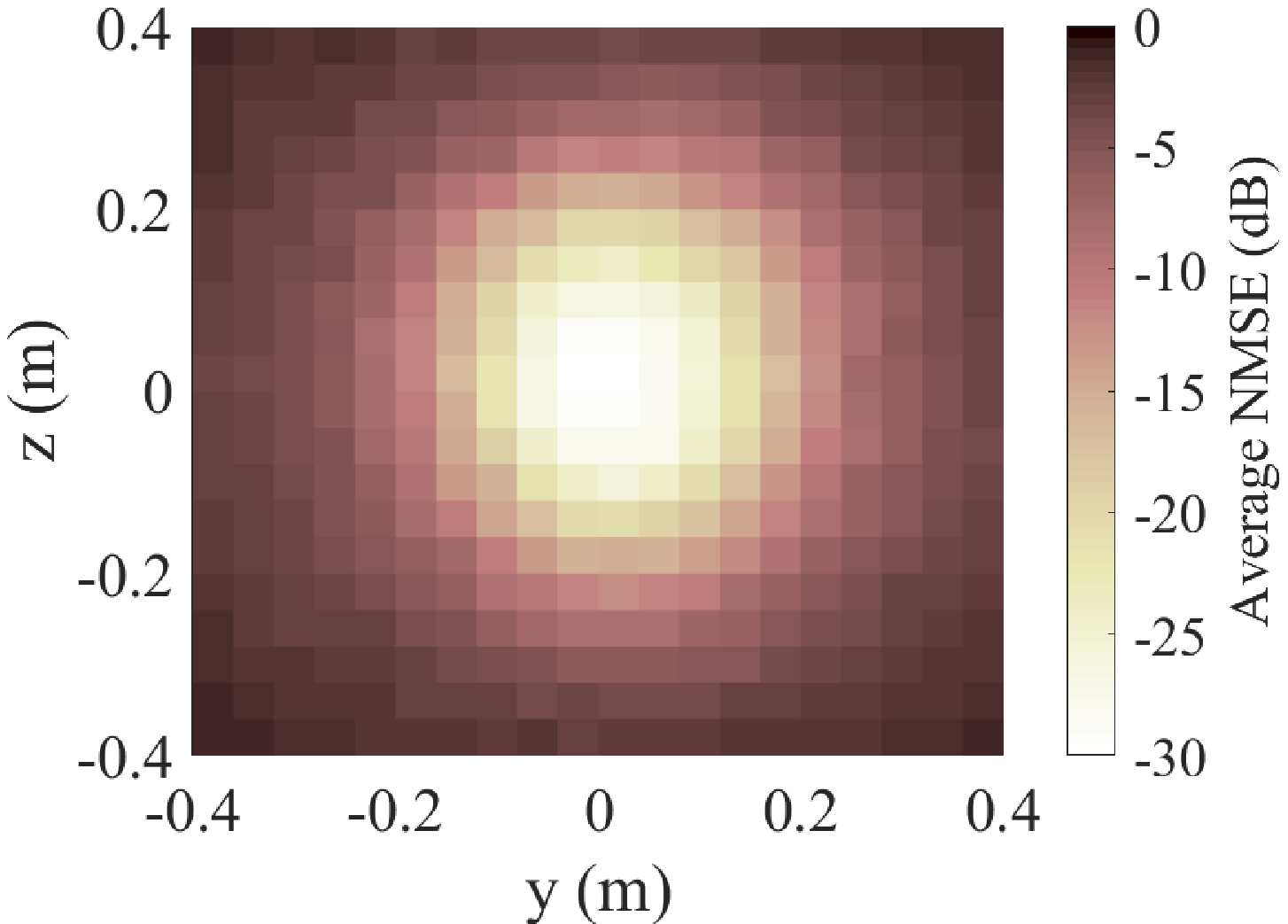}{.25\textwidth}{(b) Spherical array}\hfill}
\end{center}
\caption{Distribution of average NMSE on $y$-$z$-plane at $x=0$.}
\label{fig:ave_NMSE_yz}
\end{figure}

\begin{figure}[t!]
\figcolumn{
\fig{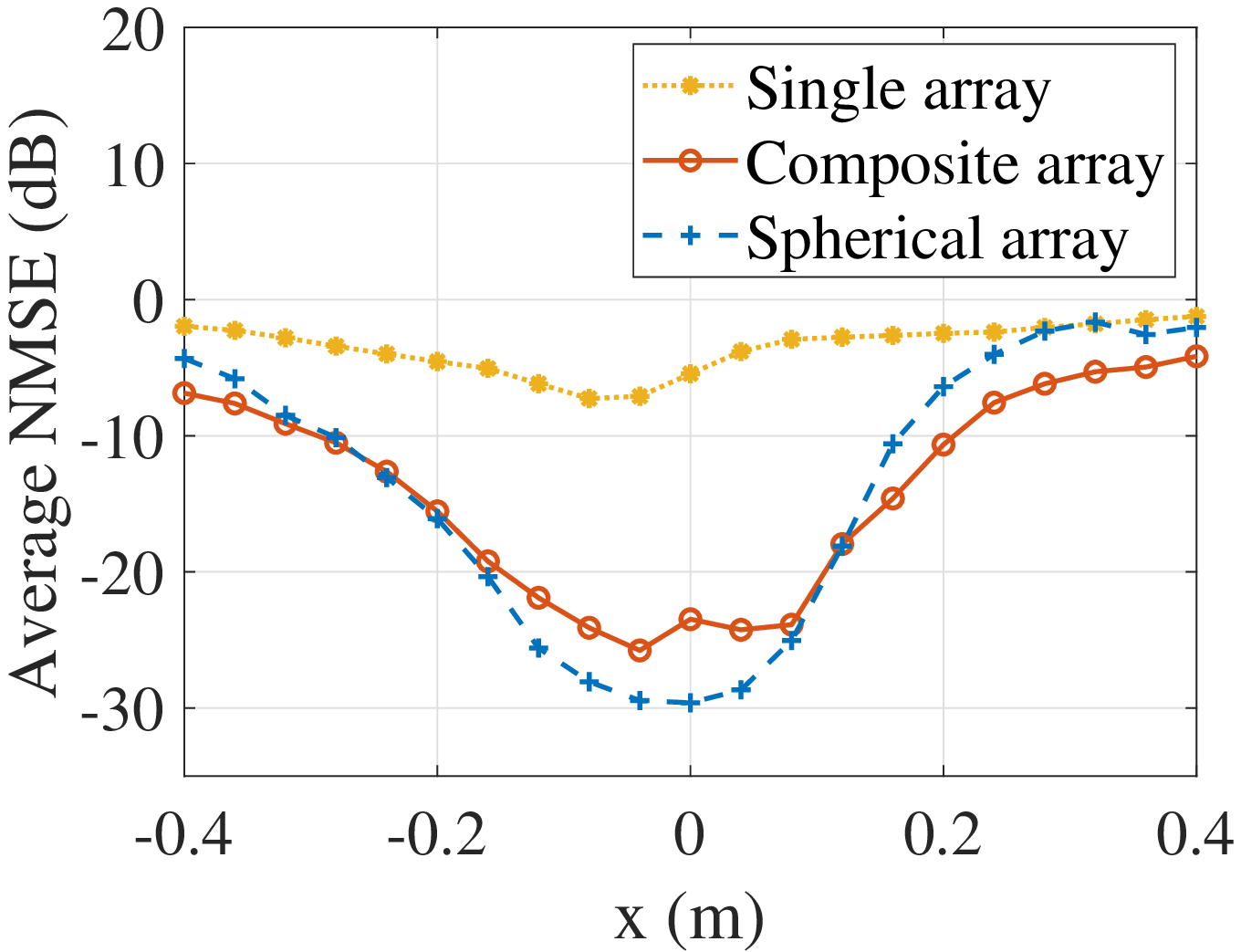}{.35\textwidth}{(a) $x$-axis}
\fig{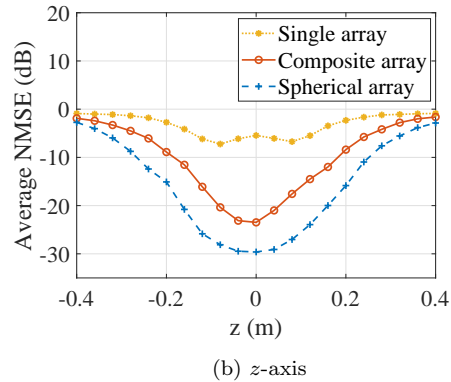}{.35\textwidth}{(b) $z$-axis}
}
\caption{Average NMSEs along the $x$- and $z$-axes.}
\label{fig:ave_NMSE_xy_axis}
\end{figure}

\begin{figure}[t!]
\figcolumn{
\fig{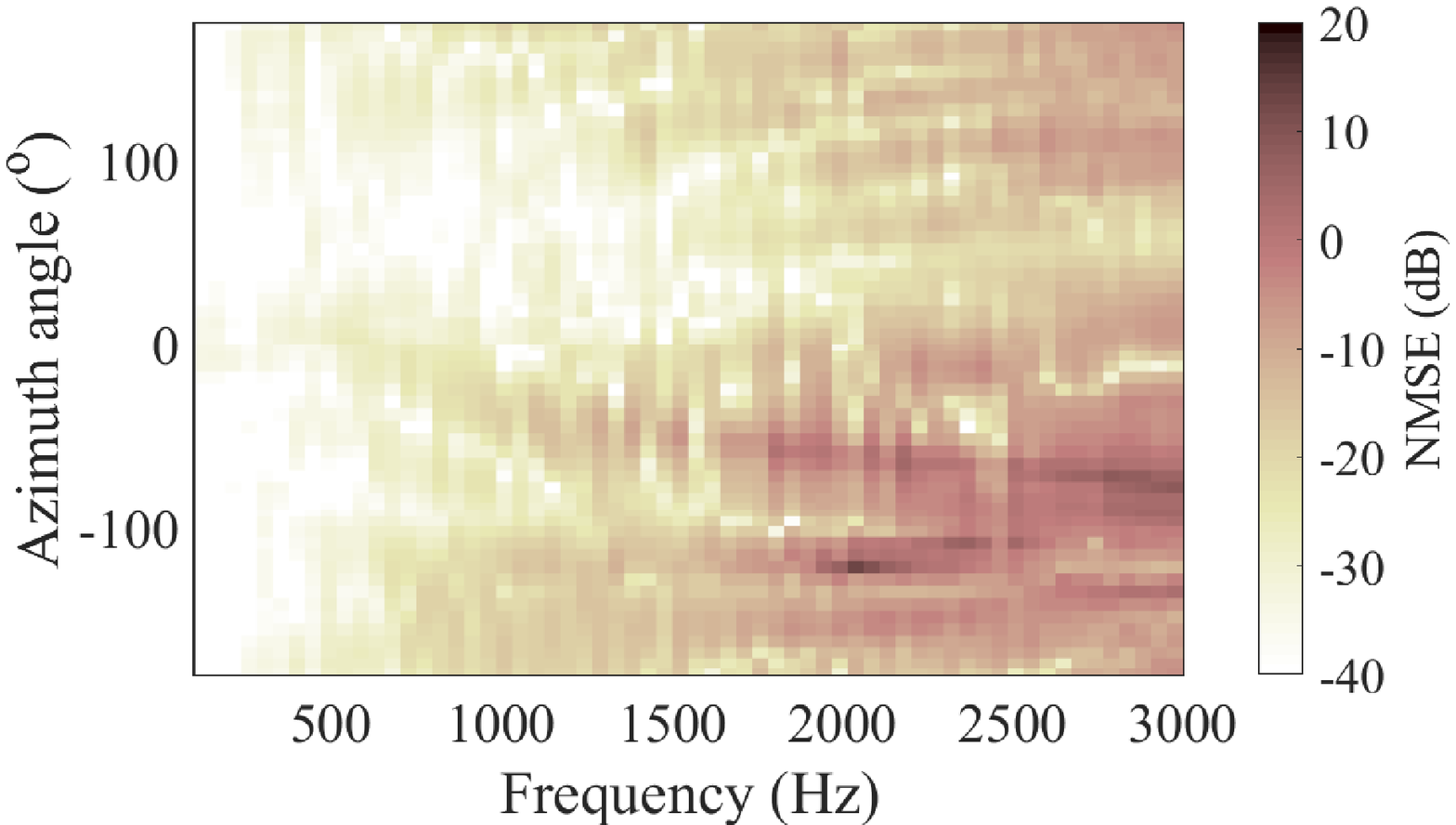}{.4\textwidth}{(a) Composite array}
\fig{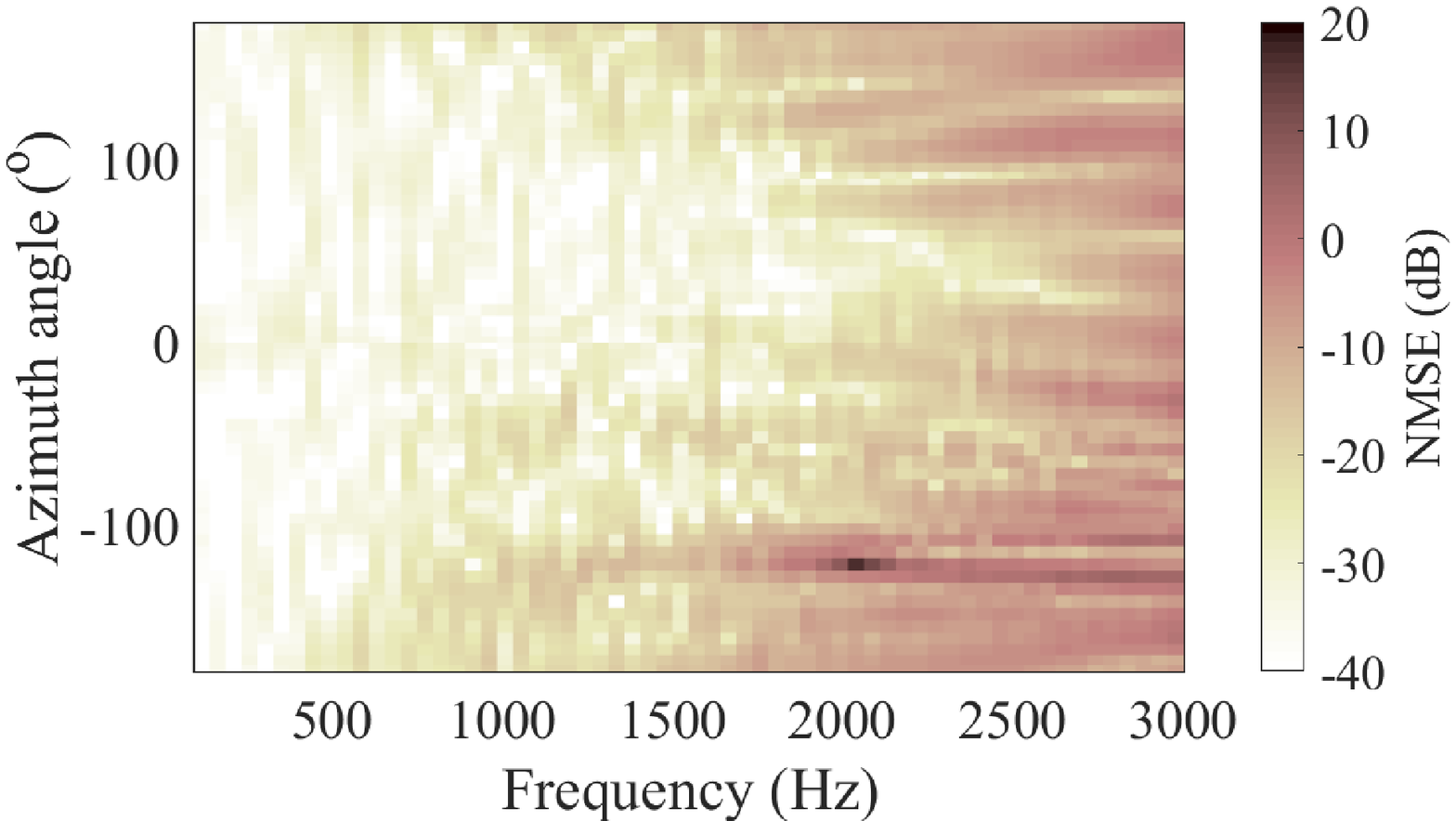}{.4\textwidth}{(b) Spherical array}
}
\caption{NMSEs at $(-0.071~\mathrm{m}, 0.071~\mathrm{m}, 0.0~\mathrm{m})$ with respect to azimuth angle and frequency.}
\label{fig:nmse_dismic}
\end{figure}

A point source was set at $(1.5~\mathrm{m}, 0.0~\mathrm{m}, 0.0~\mathrm{m})$ in the recording area. We generated binaural signals at the origin using the observed signals of the microphone arrays, changing their center positions. The positions were $21 \times 21$ grid points inside a square region of $0.8~\mathrm{m} \times 0.8~\mathrm{m}$, and 60 directions were sampled at $\theta=\pi/2$ from the directions of HRTFs. The NMSEs defined in Eq.~$\eqref{eq:NMSE}$ were averaged for directions and frequencies up to $1.6~\mathrm{kHz}$, and plotted for each position in Figs.~\ref{fig:ave_NMSE_xy} and \ref{fig:ave_NMSE_yz}. Those along the $x$- and $z$-axes are shown in Fig.~\ref{fig:ave_NMSE_xy_axis}, compared with the case of the single small array (Fig.~\ref{fig:micarray}(a)). The NMSE of the spherical array was particularly low around the origin, i.e., the listening position. The composite array achieved a comparable or lower NMSE than the spherical array in the off-center region on the $x$-$y$ plane at $z=0$. The NMSEs at $(-0.071~\mathrm{m},0.071~\mathrm{m},0.0~\mathrm{m})$ with respect to the azimuth angle and frequency are plotted in Fig.~\ref{fig:nmse_dismic}.

\section{\label{sec:3} Experiments in practical environment}

\subsection{\label{subsec:ex_in_real} Evaluation of practical system}

\begin{figure}[!t]
\begin{center}
\includegraphics[width=7cm]{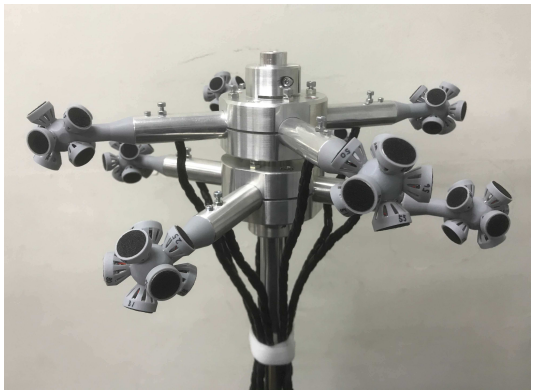}
\caption{Developed system using composite microphone array.}
\label{fig:pro_dev}
\end{center}
\end{figure}

We developed a practical composite microphone array as shown in Fig.~\ref{fig:pro_dev}. Eight small arrays consisting of eight unidirectional microphones were used, and their arrangements were identical to those in Fig.~\ref{fig:micarray}(b). The directivity pattern of several microphones was measured, and the parameter $\beta$ in Eq.~\eqref{eq:uni_dir_coef} was adapted to the measurements. We first evaluated the binaural reproduction accuracy using this microphone array. 

We reproduced the binaural signals at the origin, whose position was $1.3~\mathrm{m}$ above the ground. We placed the composite array with its center at positions A and B, and measured impulse responses of a loudspeaker at $(1.5~\mathrm{m}, 0.0~\mathrm{m}, 0.0~\mathrm{m})$ using swept-sine signals~\cite{swept_sine}. 
\begin{itemize}
 \item Position A: $(0.0~\mathrm{m}, 0.0~\mathrm{m}, 0.0~\mathrm{m})$
 \item Position B: $(-0.071~\mathrm{m}, 0.071~\mathrm{m}, 0.0~\mathrm{m})$
\end{itemize}
Again, we used the HRTF dataset of the NEUMANN KU-100 dummy head~\cite{lebedev_hrtf} for binaural rendering. We compared the reproduced binaural signals with the true signals measured using the NEUMANN KU-100 dummy head placed at the origin. We also compared the binaural signals reproduced by a single small microphone array (Fig.~\ref{fig:micarray}(a)) placed at positions A and B. The sampling frequency was set at $48~\mathrm{kHz}$.

As evaluation measures, we used the interaural time difference (ITD) and interaural level difference (ILD) defined as
\begin{align}
\mathrm{ITD} &= \argmax_{\tau} \frac{\int_{0}^{T} \bar{y}_{\mathrm{L}}(t) \bar{y}_\mathrm{R}(t+\tau) \mathrm{d}t}{\sqrt{\int_{0}^{T} \bar{y}_\mathrm{L}(t)^{2} \bar{y}_\mathrm{R}(t+\tau)^{2} \mathrm{d}t}}, \\
\mathrm{ILD} &= 10\log_{10} \frac{ \int_{0}^{T} \bar{y}_\mathrm{L}(t)^2 \mathrm{d}t}{\int_{0}^{T} \bar{y}_\mathrm{R}(t)^2\mathrm{d}t},
\end{align}
where $\bar{y}_{\mathrm{L,R}}(t)$ is the binaural signal in the time domain and $T$ is set as the signal length. These measures are important features of sound localization in humans~\cite{ITD_1600}. We removed reflections from the signals as preprocessing and applied a low-pass filter of $1.6~\mathrm{kHz}$ cutoff frequency. This is because the accurate reproduction is limited to low frequencies, as shown in the numerical simulation (Fig.~\ref{fig:nmse_dismic}), and ITD contributes to the frequency range below $1.6~\mathrm{kHz}$~\cite{ITD_1600}.

\begin{figure}[!t]
\figcolumn{\fig{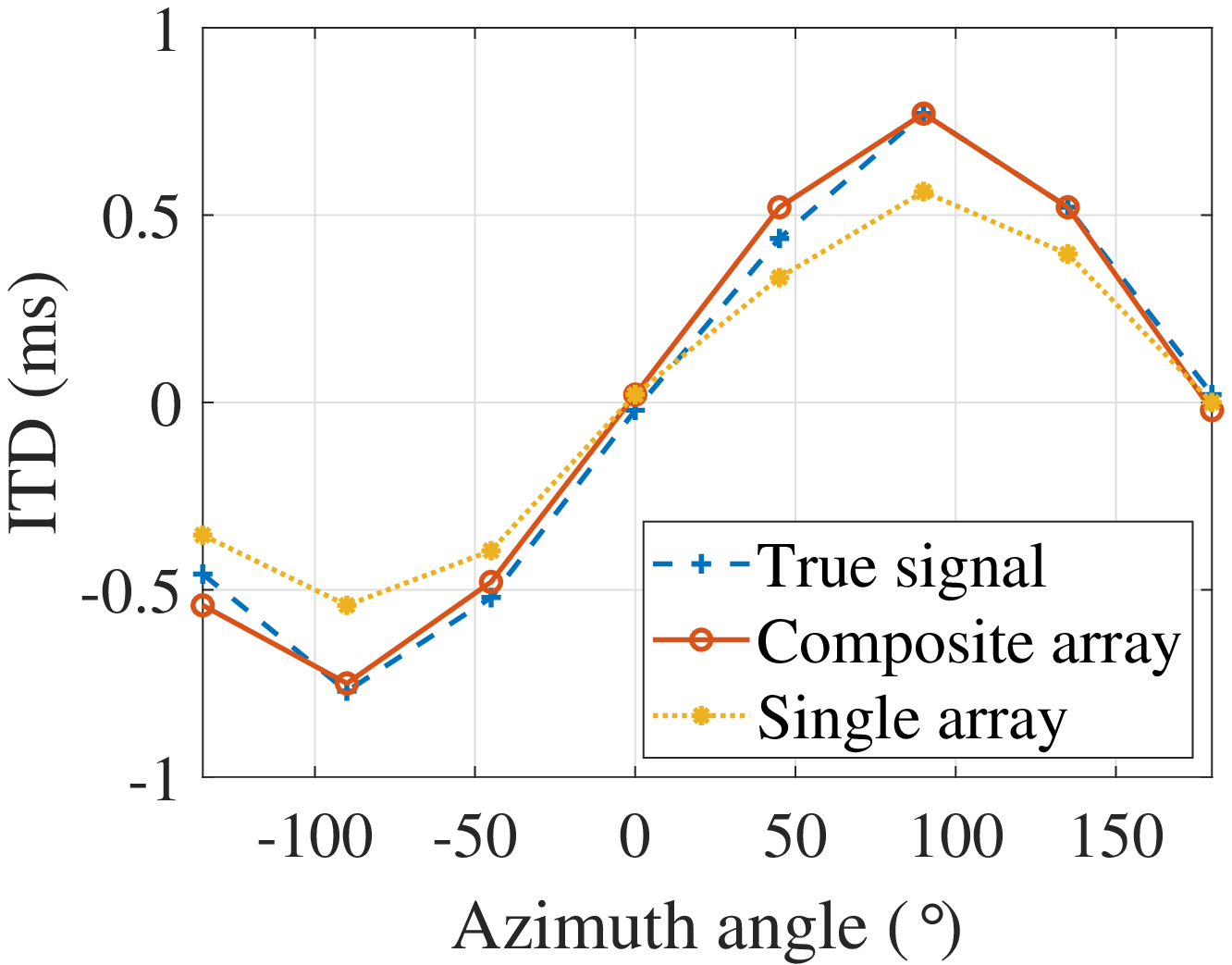}{6.5cm}{(a) Position A}
\fig{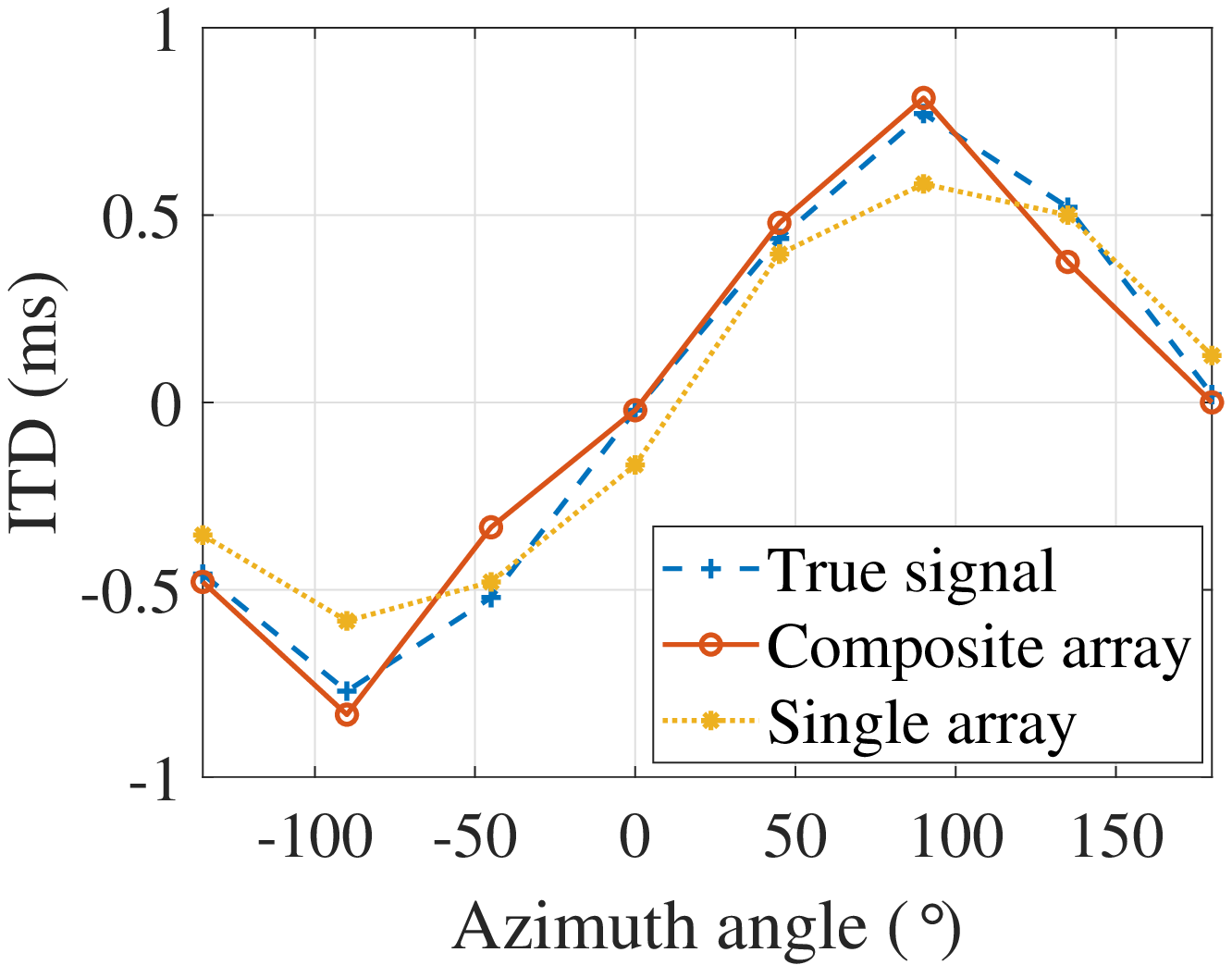}{6.5cm}{(b) Position B}}
\caption{True and reproduced ITDs at positions A and B.}
\label{fig:ITD}
\end{figure}

\begin{figure}[!t]
\figcolumn{\fig{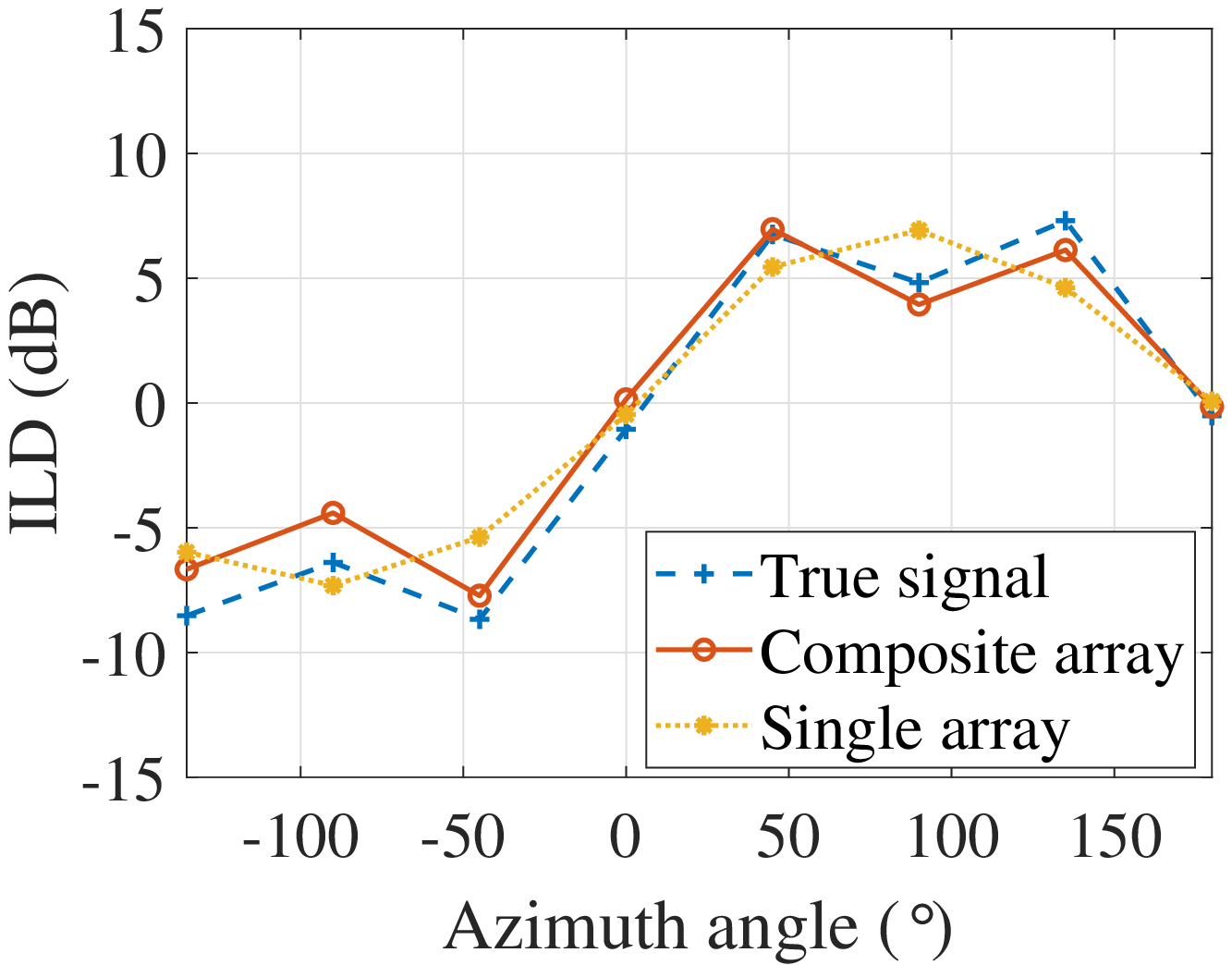}{6.5cm}{(a) Position A}
\fig{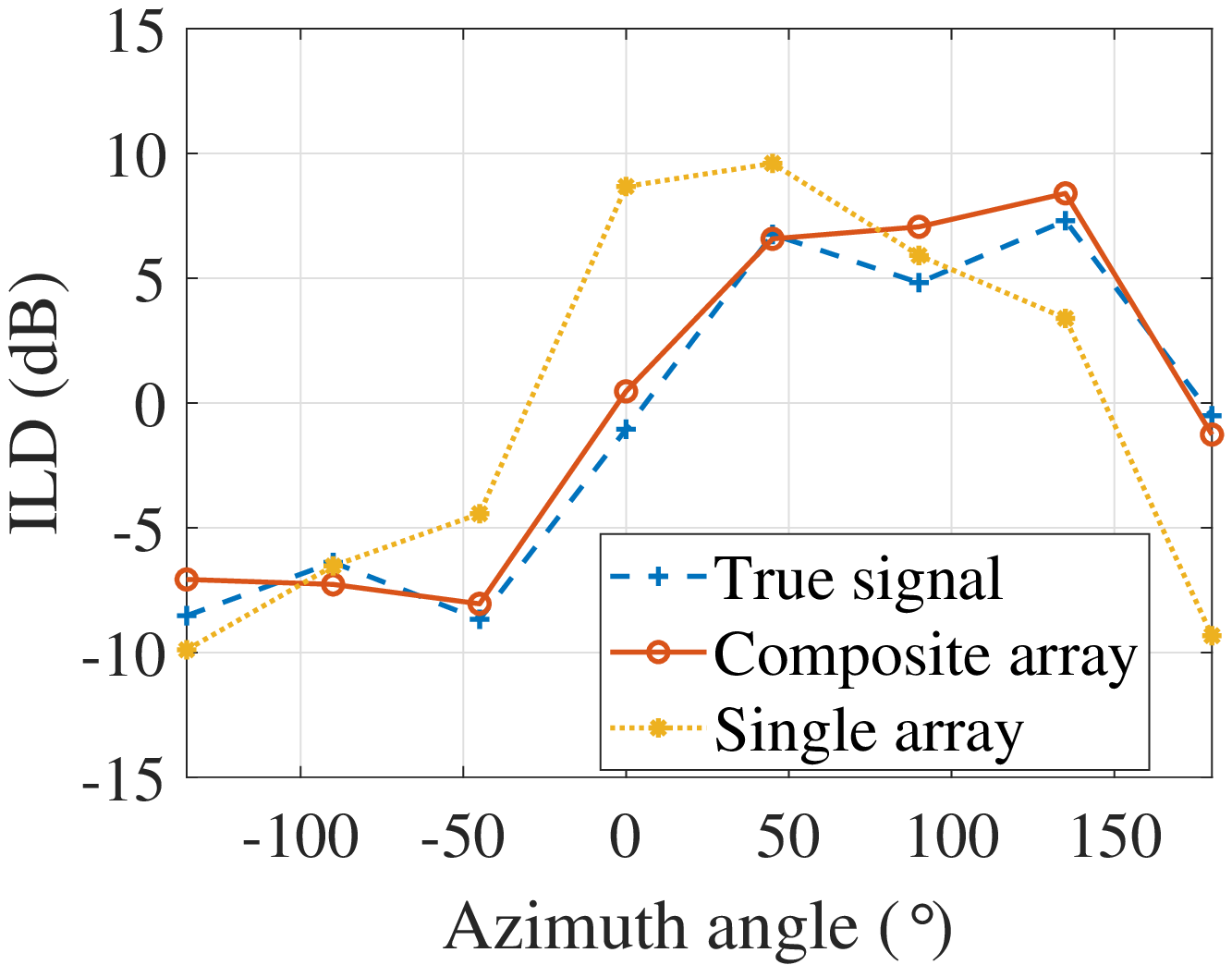}{6.5cm}{(b) Position B}}
\caption{True and reproduced ILDs at positions A and B.}
\label{fig:ILD}
\end{figure}

Figures~\ref{fig:ITD} and \ref{fig:ILD} show the ITD and ILD of the binaural signals reproduced by the proposed composite array and single small array, compared with those of the true binaural signals. The ITD and ILD are plotted  with respect to the azimuth angle at positions A and B. The ITD of both methods was close to the true one, but that of the single array was slightly smaller than the true ITD at $90^{\circ}$. The ILD at position A was accurately reproduced; however, that of the single array at position B significantly deteriorated. In the proposed composite array, the reproduced ILD at the both positions was close to the true one.

\subsection{\label{sub:listen} Listening experiments}

We evaluated our binaural reproduction method by listening experiments. Perceptual experiments using multiple stimulus with hidden reference and anchor (MUSHRA)~\cite{mushra} were conducted to compare the quality of the binaural signals generated using the composite array and single small array. Again, the HRTF dataset of the NEUMANN KU-100 dummy head was used for the binaural rendering. The sampling frequency was $48~\mathrm{kHz}$. The binaural signals at the origin were reproduced by each microphone array at positions A and B. The reference signals were the observation signals of the dummy head. As an anchor, the signals obtained by filtering the reference signals using a low-pass filter of $1.6~\mathrm{kHz}$ cutoff frequency were used. All the test signals are summarized as follows.
\begin{enumerate}
  \item Reference (hidden reference): binaural signal observed with the dummy head.
  \item C1/composite array (A): binaural signals reproduced using the composite array signals at position A
  \item C2/composite array (B): binaural signals reproduced using the composite array signals at position B
  \item C3/single array (A): binaural signals reproduced using the single small array signals at position A
  \item C4/single array (B): binaural signals reproduced using the single small array signals at position B
  \item C5/anchor: signal obtained using a low-pass filter of $1.6~\mathrm{kHz}$ cutoff frequency with respect to the reference signal
\end{enumerate}
Test participants were asked to rate the difference between the reference and test signals on a scale from $0$ to $100$. The higher score means the test signal is close to the reference signal. The test signals were repeatedly played back and the participants listened via a headphone, and the participants were able to seamlessly switch signals. Thirteen male subjects in their 20s and 30s were included, and those who scored less than 90 on the Reference more than twice were excluded from the evaluation. We used two signals, \textit{music} (acoustic guitar sound) and \textit{male speech}, for sound sources, taken from a dataset~\cite{Araki:LVAICA2021}. The binaural signals of three source directions, $\phi=45^{\circ}$, $90^{\circ}$, and $135^{\circ}$, obtained by rotational operation were included in one sequence. 

\begin{figure}[t!]
\figcolumn{
\fig{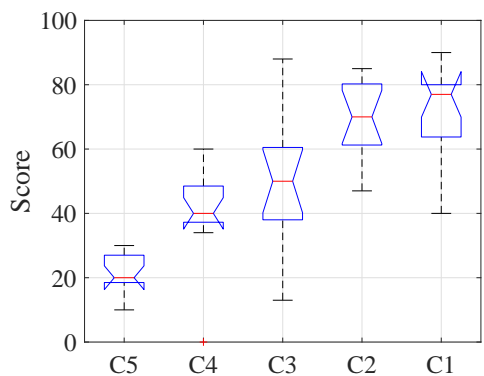}{7cm}{(a) \textit{Music}}
\fig{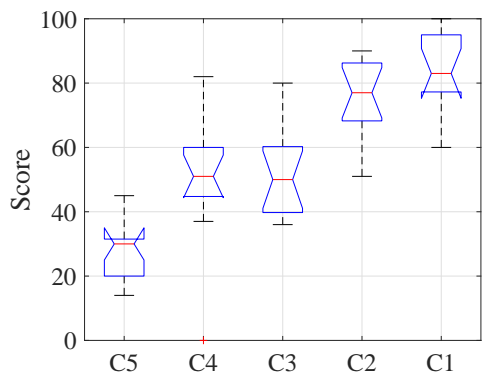}{7cm}{(b) \textit{Male speech}}
}
\caption{Box plot of listening test scores. The center bar of each box indicates the median score. The boxes and whiskers show the interquartile range (IQR) and 1.5 times IQR, respectively. The notches in the box refer to the 95\% confidence interval of the median.}
\label{fig:boxplot1}
\end{figure}

Figure~\ref{fig:boxplot1} shows the box plot of perceptual experiment scores. The median scores indicated that the proposed composite array (C1 and C2) and single array (C3 and C4) differ with 95\% confidence for both sound sources \textit{music} and \textit{male speech}. It was validated that the evaluation data followed the normal distribution by using the Lilliefors test ($p$-values $>$ 0.01). Then, the $p$-values of the Tukey--Kramer multiple comparison test were obtained as shown in Tables~\ref{tbl:box2} (\textit{music}) and \ref{tbl:box5} (\textit{male speech}). The significant difference in mean score between C1 and C3 (i.e., position A) and that between C2 and C4 (position B) were found in both sound sources. These results indicate that the perceptual quality of the binaural signals reproduced with the composite array was higher than that with the small array. Furthermore, there was no significant difference between C1 and C2; therefore, our proposed method is robust against the changes in listening position. 

\begin{table}[t!]
\caption{$p$-values (Tukey--Kramer) for \textit{music}.}
\label{tbl:box2}
\begin{ruledtabular}
  \begin{tabular}{lcccccc} 
    C5 & 1  \\
    C4 & 0.0027 & 1 \\
    C3 & 0 & 0.4183 & 1 \\
    C2 & 0 & 0 & 0.0041 & 1 \\
    C1 & 0 & 0 & 0.0016 &  0.9997 & 1 \\
    Method & C5 & C4 & C3 & C2 & C1 \\ 
  \end{tabular}
\end{ruledtabular}
\caption{$p$-values (Tukey--Kramer) for \textit{male speech}.}
\label{tbl:box5}
\begin{ruledtabular}
  \begin{tabular}{lcccccc} 
    C5 & 1  \\
    C4 & 0.0003 & 1 \\
    C3 & 0.0002 & 1 & 1 \\
    C2 & 0 & 0 & 0.0001 & 1 \\
    C1 & 0 & 0 & 0 & 0.6550 & 1 \\
    Method & C5 & C4 & C3 & C2 & C1 \\
  \end{tabular}
\end{ruledtabular}
\end{table}

\section{\label{sec:conclusion}Conclusion}

We proposed a method of binaural rendering from distributed microphone array signals. For capturing a soundfield inside a spatial region, the soundfield recording method based on harmonic analysis of infinite order is applied, which allows the use of arbitrarily placed microphones in the recording area. Typically used spherical microphone arrays are sometimes impractical because the array size should be inherently larger than the head size, and the microphone arrangement should be uniform on a sphere as much as possible.  Since our proposed method has flexibility and scalability in the microphone arrangement, we investigated a combination of small microphone arrays, which will be useful for capturing a soundfield in a large region. We also proposed a spherical-wave-decomposition-based binaural rendering method taking into consideration the distance in HRTF measurement. Compared with the conventional plane-wave-decomposition-based rendering method, the peaks and dips in the amplitude response of the binaural signals are accurately reproduced. We developed a practical composite microphone array system and evaluated the system by measuring the reproduced ITD and ILD, and conducting listening experiments. The experimental results indicated that our proposed method is robust against changes in listening position inside a range of the region in the recording area. 

\section{Acknowledgments}
We are grateful to Professor Shuichi Sakamoto for providing us with head mesh data. This work was supported by JST, PRESTO Grant Number JPMJPR18J4.





\end{document}